\documentclass[12pt,a4paper]{article}
\usepackage{amssymb}
\usepackage{amsmath}
\usepackage{color}
\usepackage[colorlinks,linkcolor=blue,citecolor=red]{hyperref}

\newcommand{\R}{\mathbb{R}}

\newcommand{\Hess}{\operatorname{Hess}}
\newcommand{\bbbr}{\mathbb R}

\newcommand\op[1]{\mathop{\rm #1}\nolimits}

\newcommand\E{{\mathcal E}}

\newtheorem{theorem}{Theorem}

\begin{document}

\title{Dispersionless integrable systems in 3D\\ and Einstein-Weyl geometry}

\author{E.V. Ferapontov and B. Kruglikov}
     \date{}
     \maketitle
     \vspace{-5mm}
\begin{center}
Department of Mathematical Sciences \\ Loughborough University \\
Loughborough, Leicestershire LE11 3TU \\ United Kingdom \\
  \ \\
  and\\
  \ \\
Institute of Mathematics and Statistics\\
NT-Faculty \\
University of Troms\o\\
Troms\o\ 90-37, Norway\\
[2ex]
e-mails: \\[1ex] \texttt{E.V.Ferapontov@lboro.ac.uk}\\
\texttt{boris.kruglikov@uit.no} \\

\end{center}

\medskip

\vspace{1cm}
\begin{abstract}

For several classes of second order dispersionless PDEs, we show that the symbols of their formal linearizations define  conformal structures which must be Einstein-Weyl in 3D (or self-dual in 4D) if and only if the PDE is integrable by the method of hydrodynamic reductions. 
This  demonstrates that  the integrability  of  these dispersionless PDEs can be seen from the geometry of their formal linearizations.

\bigskip

\bigskip

\end{abstract}

\newpage

\section{Introduction}

Let
\begin{equation}
F(x^i, u, u_{x^i}, u_{x^ix^j}, \dots)=0
\label{F=0}
\end{equation}
be a partial differential equation (PDE), where $u$ is a (scalar) function  of the independent variables $x^1, \dots, x^n$. The formal linearization of (\ref{F=0}) results upon setting $u\to u+\epsilon v$, and keeping terms of the order $\epsilon$. This leads to a linear PDE for $v$,
\begin{equation}
\ell_F(v)=0,
\label{l}
\end{equation}
where $\ell_F$ is the operator of formal linearization,
$$
\ell_F=F_u+F_{u_{x^i}}\mathcal{D}_{x^i}+F_{u_{x^ix^j}}\mathcal{D}_{x^i }\mathcal{D}_{ x^j}+\dots.
$$
(here $\mathcal{D}_{x^i}$ is the operator of total differentiation by $x^i$).
Note that, for nonlinear $F$,  the formal linearization depends on the  solution $u$.
For instance, the linearization of the dispersionless Kadomtsev-Petviashvili (dKP) equation, $u_{xt}-(uu_{x})_x-u_{yy}=0$, reads as  $v_{xt}-(uv)_{xx}-v_{yy}=0$.

The linearized equation  (\ref{l}) appears in a wide range of constructions and applications:

\begin{itemize}

\item {\it Stability analysis\/} of a  solution $u$ is based on the investigation of the spectrum of the linearized operator: this  goes back to Lyapunov.

\item {\it Symmetries\/} of the equation (\ref{F=0}) correspond to solutions of the linearized equation of the form $v=v(x^i, u, u_{x^i}, u_{x^ix^j}, \dots)$, which are required to satisfy (\ref{l}) identically modulo (\ref{F=0}): this goes back to  Lie.

\item {\it Contact invariants\/} of ordinary differential equations can be obtained from the Wilczynski invariants of linearized equations \cite{Doubrov3,Doubrov4}. 
  
\item {\it Generalized Laplace invariants\/} of PDEs appearing in the context of Darboux integrability  can be obtained from the Laplace invariants of  linearized equations 
    \cite{Zhiber, Anderson,  Juras2, Kushner, Krugl}.

\item {\it Integrability\/} of ordinary differential equations  can be seen from the structure of the differential Galois group of linearized equations: it must be Abelian, see   \cite{Morales} and references therein.

\end{itemize}

\medskip

In general, coefficients of a linear PDE have  differential-geometric meaning. In particular, its symbol can be interpreted as a symmetric tensor field. The  natural question arises: 
{\bf Can one read the integrability of a given PDE off
the geometry of its formal linearization?}
In this paper we answer this question in the affirmative  for the four particularly interesting classes of   PDEs in 3D, namely

\medskip

\noindent {\bf Equations of type I:}
\begin{equation}
\label{T1}
(a(u))_{xx} + (b(u))_{yy} + (c(u))_{tt} + 2 (p(u))_{xy} + 2 (q(u))_{xt} + 2 (r(u))_{yt} = 0,
\end{equation}
this class  was introduced in \cite{Dunajski} in the context of the `central quadric ansatz', see Remark at the end of Sect. 3. The corresponding integrability aspects were discussed in \cite{Fer5}.

\smallskip

\noindent {\bf Equations of type II:}
 \begin{equation}
f_{11} u_{xx} + f_{22} u_{yy} + f_{33} u_{tt} + 2 f_{12} u_{xy} + 2f_{13} u_{xt} + 2f_{23} u_{yt} =0,
 \label{T2}
 \end{equation}
here  the coefficients
$f_{ij}$ depend on  the first order derivatives  $u_x, u_y, u_t$ only. Equations of this type can be called quasilinear wave equations,
their integrability  was analyzed in \cite{Bur}, see also references therein.

\smallskip

\noindent {\bf Equations of type III:}
 \begin{equation}
F(u_{xx}, u_{xy}, u_{yy}, u_{xt}, u_{yt}, u_{tt})=0.
 \label{T3}
 \end{equation}
 Equations of this form are known as the dispersionless Hirota type, or Hessian type equations.
Their integrability  was studied in \cite{MaksEgor, Fer4}.

\smallskip

\noindent {\bf Equations of type IV:}
 \begin{equation}
A(u)u_x+B(u)u_y+C(u)u_t=0,
 \label{T4}
 \end{equation}
here $u$ is  a two-component column vector, and $A, B, C$ are $2\times 2$ matrices. Equations of this form are known as systems of hydrodynamic  type.
Their integrability  was investigated in \cite{FK}.
It was demonstrated in \cite{Odesskii} that coefficients of the `generic' integrable equations of the types (\ref{T2}), (\ref{T3}) and (\ref{T4}) can be parametrized by generalized hypergeometric functions.

\medskip

Equations  (\ref{T1})--(\ref{T4}) belong to the class of dispersionless PDEs. They arise  in wide range of applications in mathematical physics, general relativity, differential geometry  and the theory of integrable systems (as dispersionless limits of  integrable soliton equations of the KP/Toda type \cite{Zakharov}, see Sect. 9). In the dispersionless limit, the familiar `solitonic' integrability (based on Lax pairs,  algebro-geometric techniques, etc)  requires a modification. An adequate approach is provided by the method of hydrodynamic reductions \cite{Fer1} which is
based on the requirement of the existence of special multi-phase solutions which can be viewed as dispersionless analogues of multi-soliton/multi-gap solutions.  To make this paper as self-contained as possible, in Sect. 9 we included an Appendix with a brief overview of this approach, and further references.

\medskip

Solutions to  equations (\ref{T1})--(\ref{T4}) carry a canonical conformal structure which can be defined as follows: 
the symbol of formal linearization is a symmetric (2,0)-tensor $g^\sharp\in\Gamma(S^2TM)$
on the base manifold $M$ with coordinates $(x^1,x^2,x^3)=(x,y,t)$, which depends
on a finite jet of the  solution $u$ (in the case (\ref{T4}) we  use
the dispersion relation). This tensor is only defined up to multiplication by a non-zero factor, which makes our theory conformal. All considerations are local, and  $M$  will  be identified with an open domain of $\bbbr^3$.
We will always assume $g^\sharp$ to be non-degenerate, in this case the inverse (0,2)-tensor 
$g\in\Gamma(S^2T^*M)$ defines a metric. 
In coordinates,  $g=g_{ij}\,dx^idx^j$ where $g_{ij}$ is the inverse of the matrix 
 of $g^\sharp$. We will  assume the Lorentzian signature of $g$,  this is equivalent to the requirement of hyperbolicity of the corresponding PDE (see Sect. 9 for a  discussion of the elliptic case).
The conformal class $[g]$ of the metric $g$ is the key invariant responsible for the linearizability/integrability of the equations under study.
Our main results can be summarized as follows:

\begin{itemize}

\item Equations (\ref{T1})--(\ref{T4}) are linearizable (by a transformation from the natural equivalence group specified in each particular case below) if and only if  the corresponding conformal structures $g$ are conformally flat on every solution (an extra condition is needed in the case (\ref{T4})). This provides a simple  linearizability test based on the vanishing of the corresponding Cotton tensor.

\item Equations (\ref{T1})--(\ref{T4}) are integrable by the method of hydrodynamic reductions  if and only if  the corresponding conformal structures $g$ are Einstein-Weyl on every solution (again an extra condition is needed in the case (\ref{T4})). Recall that an Einstein-Weyl structure consists of a symmetric connection $\mathbb{D}$ and a conformal structure $g$  such that

\noindent (a) the connection $\mathbb {D}$ preserves the conformal class: $\mathbb{D}[g]=0$, 

\noindent (b) the trace-free part of the symmetrized Ricci tensor of $\mathbb {D}$ vanishes. 

In coordinates, this gives
\begin{equation}
\mathbb{D}_kg_{ij}=\omega_k g_{ij}, ~~~ R_{(ij)}=\Lambda g_{ij},
\label{EW}
\end{equation}
where $\omega=\omega_kdx^k$ is a covector, $R_{(ij)}$ is the symmetrized Ricci tensor of $\mathbb{D}$,  and $\Lambda$ is some function \cite{Cartan}.  In fact one needs to specify $g$ and $\omega$ only, then the first set of equations uniquely defines $\mathbb{D}$. We point out that for all examples considered in this paper, the covector $\omega$ is expressed in terms of  $g$ by the  universal explicit formula 
\begin{equation}
\omega_k=2g_{kj}\mathcal{D}_{x^s}(g^{js})+\mathcal{D}_{x^k}(\ln\det g_{ij}).
\label{omega}
\end{equation}
Note that in 3D this formula  is  invariant
under the transformation $g\to \lambda g, \ \omega \to \omega + d\ln\lambda$,
which is characteristic of the Einstein-Weyl geometry. 

\end{itemize}



We recall that the Einstein-Weyl equations  (\ref{EW}) are  integrable by   twistor-theoretic methods \cite{Hitchin}. Thus, solutions of integrable PDEs carry `integrable' geometry.  Equivalently, one can say that  second order dispersionless integrable systems in 3D (having non-degenerate  symbol) can be viewed as reductions of the Einstein-Weyl conditions, which therefore play the role of  a universal `master-equation'
\cite{Ward1}. 
Let us mention that  relations of dispersionless integrable systems to the 
Einstein-Weyl geometry have been discussed in \cite{Ward, Calderbank, Calderbank1, LeBrun, Dun4, Dun6, Dun7, Dun3, God}.

Given a class of integrable PDEs such as (\ref{T2})--(\ref{T4}), the verification of the Einstein-Weyl conditions  (\ref{EW}) can be a formidable task, primarily due to a rather intricate structure of the integrability conditions. A way to bypass computational difficulties is to use the result of Cartan \cite{Cartan} which says that the Einstein-Weyl property of a triple  $(\mathbb{D}, \ g, \ \omega)$ is equivalent to the existence of a two-parameter family of surfaces which are null with respect to the conformal structure $g$ (that is, tangential to the null cones of $g$), and totally geodesic in the Weyl connection  $\mathbb{D}$. In the context of dispersionless integrable systems, such surfaces are provided by the corresponding dispersionless Lax pairs: these consist  of  $\lambda$-dependent vector fields $X, Y$ which are required to commute modulo the equation,  identically in the `spectral parameter' $\lambda$ (for all classes of PDEs discussed in this paper,  the existence of such Lax pairs is equivalent to the integrability by the method of hydrodynamic reductions). Note that $X$ and $Y$ may contain derivatives with respect to ${\lambda}$. Taking integral surfaces of the distribution spanned by $X, Y$ in the extended four-space with coordinates $x, y, t, \lambda$, and projecting them down to the space of independent variables $x, y, t$, we obtain the required two-parameter family of null totally geodesic surfaces.
Computationally, this approach has an advantage, allowing one to avoid working with the full set of integrability conditions
(the problem is to prove  the existence of a Lax pair, and this is where the integrability conditions are needed). 
In this approach, the key object within the triple  $(\mathbb{D}, \ g, \ \omega)$ is the connection  $\mathbb{D}$, which is uniquely specified by the given two-parameter family of null totally geodesic surfaces. Both Einstein-Weyl conditions (\ref{EW}) will be satisfied automatically. 



\medskip

Our main results relating linearizability/integrability to geometry of formal linearizations are proved in Sect. 3-6 (we find it more convenient to treat the above four classes separately: explicit forms of the corresponding linearizability/integrability conditions are rather different). Known integrable equations of types  (\ref{T1})--(\ref{T4}) provide an abundance of   Einstein-Weyl structures parametrized by elementary functions, elliptic functions, modular forms and Painlev\'e transcendents.  Some further examples are collected in Sect. 7.
Most of the examples of Einstein-Weyl structures exhibited in this paper are apparently new (otherwise, a reference is given).

In Sect. 8 we discuss geometric aspects of  integrability of second order dispersionless PDEs in 4D, indicating  that the associated conformal structures must be self-dual. This is in agreement with the fundamental fact that the Einstein-Weyl equations are reductions of the equations of self-duality  \cite{ Jones, Calderbank, Calderbank1}. 

The method of hydrodynamic reductions, which provides an efficient approach to the integrability of equations   (\ref{T1})--(\ref{T4}), is summarized in Sect. 9.

In  calculations of the Cotton tensor and the Einstein-Weyl conditions we used  symbolic packages of Maple. We shall omit  unnecessary lengthy formulae from the text. All relevant programs, including the integrability conditions of equations  (\ref{T1})--(\ref{T4}), and details of proofs from Sect. 3-6, are available from arXiv:1208.2728v3. 

 
 \bigskip
 
\noindent  {\bf Conventions.} All our considerations are micro-local, i.e.
local for a solution, with the size of neighbourhood also depending on  (jet of) the solution. 
We work either in the real smooth category, or in the complex holomorphic category. 
In the latter case we only assume non-degeneracy of the symbol, while in the former 
we assume that the symbol is hyperbolic
(see the next section for precise definition). 
As our approach to the dispersionless integrability is based on the method of hydrodynamic
reductions, which generally refers to hyperbolic systems,  we assume the Lorentzian signature of g
in 3D, and the neutral signature in 4D, but this requirement 
can be removed without restricting the generality if the PDE in question is analytic.
In fact one can treat the elliptic case by the complexification approach, 
see Sect. 9 for more details.

Although results of this paper are local, we perceive that global versions may be available through the twistor theory. 
 
 \section{Preliminaries}

In this section we discuss the necessary background material.
For simplicity we restrict to the case of a single partial differential equation $\E$ of the second order. 
A scalar  second order  differential operator is a function $F\in C^\infty(J^2M)$ on the space 
of 2-jets of the base manifold $M$, and the equation $\E=\{F=0\}$ is a submanifold in $J^2M$.

As described in the Introduction, the linearization operator $\ell_F$ is a second order linear differential
operator defined modulo the equation $\E$, which means that its coefficients, being functions of the finite order jets 
of $u$, are subject to this PDE. 
The important property of formal linearization is its contact invariance, which
ensures that  contact transformations lift naturally to the 
tangent bundle (covering) of the equation.
To the best of our knowledge, this invariance was explored for the first time in \cite{Anderson}.

In this paper we need a simpler fact that the symbol $\sigma_F$ of the linearization operator $\ell_F$,
also called the symbol of $F$, is contact invariant.
The symbol is a bi-vector $\sigma_F\in C^\infty(\E,S^2TM)$ depending on the 2-jet
$[u]_x^2\in\E$; in local coordinates, $\sigma_F=F_{u_{x^ix^j}}\partial_{x^i}\partial_{x^j}$.
Let us briefly indicate the proof.

A contact transformation $\Phi:J^1M\to J^1M$ lifts naturally to a transformation
$\Phi^{(1)}:J^2M\to J^2M$ (the latter is usually defined on an open dense subset of $J^2M$ due to mixing
of dependent and independent variables). The fibers of the projection $\pi_{2,1}:J^2M\to J^1M$
are affine spaces associated to the fibers of $S^2T^*M$. The prolongation $\Phi^{(1)}$ is fiberwise projective on them. Its symbol $\sigma_{\Phi^{(1)}}:S^2T^*M\to S^2T^*M$ (here and below the bundles
are pulled back to $\E$ via natural projections) is the differential of this projective transformation.
Using the properties of the symbol, cf. \cite{Kras,Spencer}, we conclude
 $$
\sigma_{(\Phi^{(1)})^*(F)}=\sigma_{F\circ\Phi^{(1)}}=\sigma_F\circ\sigma_{\Phi^{(1)}},
 $$
This is the required contact invariance of the symbol $\sigma_F:S^2T^*M\to\R$ .

A non-zero covector $p\in T^*_xM$ is called characteristic for $F$ at $[u]_x^2\in\E$ if
$\sigma_F([u]_x^2)(p,p)=0$. The projectivized (complexified) set of characteristic covectors is 
called the (complex) characteristic variety $\op{Char}(\E)$ at $[u]_x^2$. The equation $\E$ 
is hyperbolic if its complex characteristic variety $\op{Char}(\E)$ is the complexification 
of a real variety. The characteristic variety is invariant under contact transformations. 
Indeed, let $g_\E=\op{Ker}(\sigma_F)=T\E\cap\op{Ker}d\pi_{2,1}\subset S^2T^*M$ be the symbol of $\E$.
Then the claim follows from the  fact \cite{Spencer} that a covector $p$
is characteristic whenever $p^2=p\cdot p\in g_\E$.



The basic object of our study is the bi-vector  $\sigma_F=g^\sharp$, and when this bi-vector is non-degenerate,
we consider the dual (conformal) metric $g\in S^2T_x^*M$ which depends only on $[u]_x^2\in\E$ (we point out that the projectivized null cone of $g$ is dual to the  characteristic variety).
Together with the 1-form $\omega$, the metric $g$ uniquely defines  the connection $\mathbb{D}$ by the first equation
of (\ref{EW}), and the triple $(\mathbb{D}, \ g, \ \omega)$
defines an Einstein-Weyl structure if the second equation of (\ref{EW})
is satisfied. 

We say that a certain tensor (Cotton, Einstein-Weyl, etc), which depends on higher order jets of  $u$, vanishes on every solution, if it vanishes modulo the equation $\E$, meaning again that 
jets  $u$ are constrained by the equation, and a finite number of its differential consequences.
In practice we eliminate, say, all higher order derivatives of $u$ containing differentiation by $t$ 
more than once (i.e. $u_{tt}, u_{ttt}, u_{xtt}$, etc), 
and equate to zero terms at the remaining higher order derivatives.
Since all objects depend  on a finite order jet of $u$,  we do not rely upon smooth solvability of  
the equation, and  carry out calculations formally using the geometric theory of PDEs.

\medskip

\noindent{\bf Proposition}
{\it For an equation $\E$, the properties for the conformal  metric $g$ to have the Cotton tensor zero on every solution, or  to satisfy the Einstein-Weyl conditions 
on every solution, are contact invariant.}

\medskip
 
\centerline{\bf Proof:}
Let $F\circ\Phi^{(1)}=\tilde F$ be the transformed operator. Then  the prolonged contact
transformation maps $\sigma_{\tilde F}$ to $\sigma_F$, and consequently the metric $g$, is mapped to 
the corresponding metric $\tilde g$. Denote $\tilde\omega$ the pull-back of $\omega$
(higher prolongations of $\Phi$ are used at this step). The contact map $\Phi$ sends a solution
$\tilde S$ of $\tilde\E=\{\tilde F=0\}$, considered as a Legendrian submanifold in $J^1M$, to 
a solution $S$ of $\E$, and the vanishing of the Cotton tensor of $\tilde g$ on $\tilde S$ is equivalent 
to the same condition for $g$ on $S$. Similarly, the Einstein-Weyl property for
$(\tilde g,\tilde\omega)$ on $\tilde S$ is equivalent to the same property for
$(g,\omega)$ on $S$. 

Classical solutions (projecting diffeomorphically onto $M$) may  be mapped to multi-valued solutions
(Legendrian submanifolds), however, locally most of them are mapped to  classical solutions.
More precisely, if we consider $(k+1)$-jets of solutions, then the prolongation $\Phi^{(k)}$ is defined
on an open dense subset thereof. Thus, the vanishing of the Cotton tensor, and the trace-freeness for the
symmetrized Ricci tensor on every solution of $\E$, implies similar properties for almost all solutions of $\tilde\E$.
The latter we quantify to hold on an open dense set of the prolonged equation 
$\tilde\E{}^{(k-1)}\subset J^{k+1}M$, and this by continuity 
 implies the required property for all solutions. The Proposition is proved.

\medskip

Thus we obtain a covariant approach to integrability. Notice however that the integrability 
by the method of hydrodynamic reductions, as well as the explicit form  
(\ref{omega}) of the covector $\omega$, are coordinate-dependent. More about this will be said in  Concluding Remarks.

The above discussion covers  integrable PDE of types (\ref{T1})--(\ref{T3}). For 
systems of first order PDEs, such as (\ref{T4}), the theory is similar:  the only
difference is that, by virtue of the Lie-B\"acklund theorem,  contact transformations  should
be changed to  point transformations, i.e. diffeomorphisms of $J^0(M,\R^2)=M\times\R^2$.

\section{Equations of type I}

In this section we consider equations of the form (\ref{T1}),
$$
(a(u))_{xx} + (b(u))_{yy} + (c(u))_{tt} + 2 (p(u))_{xy} + 2 (q(u))_{xt} + 2 (r(u))_{yt} = 0.
$$
Their integrability was investigated in \cite{Fer5} based on the method of hydrodynamic reductions. This boils down to the requirement of the existence of an infinity of multi-phase solutions, which imposes strong constraints on the coefficients of the equation, and provides an efficient classification criterion (see Sect. 9 for a brief summary of the method). 
To formulate the classification  result we introduce the symmetric matrix
$$
V(u)=\left(\begin{array}{ccc}
a'&p'&q'\\
p'&b'&r'\\
q'&r'&c'
\end{array}
\right),
$$
where prime denotes differentiation by $u$. The classification is  performed modulo (complex) linear changes of the independent variables $x, y, t$, as well as transformations $u\to \varphi(u)$, which constitute the equivalence group of our problem.

 \begin{theorem}\cite{Fer5} \label{Thm1}
Equation (\ref{T1}) is integrable by the method of hydrodynamic reductions if and only if the matrix $V(u)$ satisfies the constraint
 \begin{equation}
V''=(\ln \det V)' V'+kV,
\label{b}
 \end{equation}
for some scalar function k.
Modulo equivalence transformations, this leads to the five canonical forms of nonlinear integrable models:
 $$
u_{xx}+u_{yy}-(\ln(e^u-1))_{yy}-(\ln(e^u-1))_{tt}=0,
 $$
 $$
u_{xx}+u_{yy}-(e^u)_{tt}=0,
 $$
 $$
(e^u-u)_{xx}+2u_{xy}+(e^u)_{tt}=0,
 $$
 $$
u_{xt}-(uu_x)_x-u_{yy}=0,
 $$
 $$
(u^2)_{xy}+u_{yy}+2u_{xt}=0.
 $$
Examples 2 and 4 are the familiar Boyer-Finley (BF) and the dKP equations.
 \end{theorem}

We point out that the constraint (\ref{b}), which  implies  $V''\in span \{V, V'\}$,  means that the `curve' $V(u)$ lies in a two-dimensional linear subspace of the space of $3\times 3$ symmetric matrices.  The classification of normal forms of such linear subspaces, which is equivalent to the classification of pencils of conics, leads to the five canonical forms of Theorem \ref{Thm1}. It was pointed out  by D. Calderbank that equations of the form (\ref{T1}) are related to generalized Nahm equations with the gauge group $SDiff(\Sigma^2)$. In this language, the five canonical forms of Theorem \ref{Thm1} correspond to the five types of generalized Nahm equations obtained in \cite{Calderbank}. For the dKP equation this correspondence was explicitly demonstrated in \cite{Dunajski}.

The  linearized equation (\ref{T1}) is
$$
a'(u)v_{xx} + b'(u)v_{yy} + c'(u)v_{tt} + 2 p'(u)v_{xy} + 2 q'(u)v_{xt} + 2 r'(u)v_{yt} +\dots = 0,
$$
where dots denote terms with lower order derivatives of $v$. Its symbol defines a conformal structure $g=g_{ij}(u)dx^idx^j$ where $(x^1, x^2, x^3)=(x, y, t)$, and the matrix $g_{ij}$ is the inverse of $V$.  Our first result is as follows.

 \begin{theorem}\label{Thm2}
Equation (\ref{T1}) is linearizable by a transformation from the equivalence group 
if and only if the conformal structure $g$ is conformally flat on every solution.
 \end{theorem}

\centerline{\bf Proof:}

The condition responsible for conformal flatness in three dimensions is the vanishing of the Cotton tensor,
\begin{equation}
\nabla_r(R_{pq}-\frac{1}{4}Rg_{pq})=\nabla_q(R_{pr}-\frac{1}{4}Rg_{pr}),
\label{cot}
\end{equation}
where $R_{pq}$ is the Ricci tensor, $R$ is the scalar curvature, and $\nabla$ denotes covariant differentiation in the Levi-Civita connection of   $g$. Calculating (\ref{cot}) and using  (\ref{T1}) and its differential consequences to eliminate all higher order partial derivatives of $u$ containing differentiation by $t$ more than once, we obtain expressions which have to vanish identically
in the remaining higher order derivatives of $u$ (without  loss of generality we will assume that $c(u)=u$, this can  be achieved by a transformation from the equivalence group). Requiring the vanishing of coefficients at the remaining  derivatives of $u$, we obtain that all entries of the matrix $V$ must be constant, which leads to linear equations. 
This finishes the proof of Theorem \ref{Thm2}.

\medskip

It turns out that  conformal structures corresponding to all five integrable  models from Theorem \ref{Thm1} satisfy the Einstein-Weyl property. In fact, this follows from the construction  of \cite{Calderbank} which provides Einstein-Weyl structures from solutions of the gauge field equations 
with the gauge group $SDiff(\Sigma^2)$ modelled on Riccati spaces; our goal here is to present the
explicit formulae. In addition, for all equations from Theorem \ref{Thm1} we present dispersionless Lax pairs in the form $[X, Y]=0$ where $X$ and $ Y$ are $\lambda$-dependent vector fields which commute modulo the equation. Projecting integral surfaces of the distribution spanned by  $X$ and $Y$ from the extended space of coordinates $x, y, t, \lambda$ down to the space of  independent variables $x, y, t$, we obtain  two-parameter families of null totally geodesic surfaces of the corresponding Einstein-Weyl structures.

\bigskip

\noindent {\bf Equation 1:} $u_{xx}+u_{yy}-(\ln(e^u-1))_{yy}-(\ln(e^u-1))_{tt}=0.$

\noindent Conformal structure\footnote{Here and in what follows we choose the `simplest' representative metric $g$ within the conformal class $[g]$. In all cases it is given by either the inverse or the cofactor matrix of $g^\sharp$.}: $g=dx^2+(1-e^u)dy^2+(e^{-u}-1)dt^2$.

\noindent Covector: $\omega=\frac{e^{u}+1}{e^{u}-1}u_xdx-u_ydy+u_tdt$.

\noindent Lax pair:
$$
\begin{array}{c}
X=\partial_y+\sqrt{e^u-1}\sin \varphi  \ \partial_x+\left(\frac{e^u}{e^u-1}u_t-\frac{\cos \varphi }{\sqrt{1-e^{-u}} } u_x\right) \partial_{\lambda}, \\
\ \\
Y=\partial_t + \sqrt{1-e^{-u}}\cos \varphi \  \partial_x +\left(\frac{1}{1-e^u}u_y+\frac{\sin\varphi}{\sqrt{e^u-1}}u_x  \right) \partial_{\lambda},
\end{array}
$$
here $\varphi=- \arctan (e^{-u/2}\tan \lambda /2).$

\medskip

\noindent {\bf Equation 2:} $u_{xx}+u_{yy}-(e^u)_{tt}=0$ (BF equation).

\noindent Conformal structure: $g=dx^2+dy^2-e^{-u}dt^2$.

\noindent Covector: $\omega=-u_xdx-u_ydy+u_tdt$.

\noindent This Einstein-Weyl structure was  obtained in \cite{Ward}, see also \cite{LeBrun}.

\noindent Lax pair:
$$
X=\partial_y-e^{u/2}\sin \lambda \ \partial_t -\frac{1}{2}(u_x+e^{u/2}u_t \cos \lambda)\partial_{\lambda}, ~~~
Y=\partial_x-e^{u/2}\cos \lambda \ \partial_t +\frac{1}{2}(u_y+e^{u/2}u_t \sin \lambda)\partial_{\lambda}. 
$$

\noindent {\bf Equation 3:} $(e^u-u)_{xx}+2u_{xy}+(e^u)_{tt}=0.$

\noindent Conformal structure: $g=2dxdy+(1-e^u)dy^2+e^{-u}dt^2$.

\noindent Covector: $\omega=-u_xdx+(2e^uu_x-u_y)dy+u_tdt$.

\noindent Lax pair:
$$
X=\partial_t-\lambda \partial_x +(\lambda^2+1)u_x\partial_{\lambda}, ~~~
Y=\partial_y+\frac{1}{2}(e^u(\lambda^2+1)-1) \partial_x -\frac{1}{2}e^u(u_t+\lambda  u_x)(\lambda^2+1)\partial_{\lambda}. 
$$

\noindent {\bf Equation 4:} $u_{xt}-(uu_x)_x-u_{yy}=0$ (dKP equation).

\noindent Conformal structure: $g=4dxdt-dy^2+4udt^2$.

\noindent Covector: $\omega=-4u_xdt$.

\noindent This Einstein-Weyl structure was obtained in \cite{Dun4}.

\noindent Lax pair:
$$
X=\partial_y-\lambda \partial_x +u_x\partial_{\lambda}, ~~~
Y=\partial_t-(\lambda^2+u) \partial_x +(u_x\lambda +u_y)\partial_{\lambda}. 
$$

\noindent {\bf Equation 5:} $(u^2)_{xy}+u_{yy}+2u_{xt}=0.$

\noindent Conformal structure: $g=2dxdt+dy^2-2udydt+u^2dt^2$.

\noindent Covector: $\omega=2u_xdy+2(u_y-uu_x)dt$.

\noindent Lax pair:
$$
X=\partial_y-{\lambda} \partial_x +2u_x\lambda \partial_{\lambda}, ~~~
Y=\partial_t+(\frac{1}{2}{\lambda}^2+u{\lambda}) \partial_x -(u_x{\lambda} +u_y+2uu_x)\lambda \partial_{\lambda}. 
$$

All of the above conformal structures $g$ and covectors $\omega$  can be represented  in terms of the matrix $V(u)$ as follows:
\begin{equation}
g=(dx\ dy\ dt) V^{-1}
\left(
\begin{array}{c}
dx \\
dy\\
 dt
 \end{array}\right), ~~~
 \omega=2(dx\ dy\ dt) V^{-1}V'
\left(
\begin{array}{c}
u_x \\
u_y\\
 u_t
 \end{array}\right)-d(\ln\det V).
\label{g}
\end{equation}
They satisfy the Einstein-Weyl equations (\ref{EW}) if and only if  $V(u)$ satisfies the integrability condition (\ref{b}).
Setting $(x, y, t)=(x^1, x^2, x^3)$ one can represent the components of $\omega=\omega_kdx^k$ by the formula (\ref{omega}). 
It turns out that exactly the same formula holds for all other classes of dispersionless PDEs discussed in this paper.  The covector $\omega$ is related to the symbol of formal linearization via the  identity
$$
g^{ij}v_{x^ix^j}=\nabla^i\nabla_iv-\frac{1}{2}\omega^i\nabla_iv
$$
where $\nabla^i=g^{ik}\nabla_k, \ \omega^i=g^{ik}\omega_k$, and $\nabla$ denotes covariant differentiation in the Levi-Civita connection of the metric $g_{ij}$. Note that the right hand side of this identity can be interpreted as a special case of the  Weyl wave operator of weight zero \cite{Dun6}. The second main result of this section is


\medskip

 \begin{theorem}\label{Thm3}
Equation (\ref{T1}) is integrable by the method of hydrodynamic reductions if and only if the corresponding  conformal structure $g$ is Einstein-Weyl on every solution, with the covector $\omega$ given by (\ref{omega}).
 \end{theorem}
 
\bigskip

\centerline{\bf Proof:}

\medskip

Given an equation of the form (\ref{T1}), the conformal structure of its formal linearization is  
$g=(dx\ dy\ dt)V^{-1}(dx\ dy\ dt)^t$. We will seek a covector $\omega$ in the form
$\omega=(dx\ dy\ dt)T(u_x\ u_y \ u_t)^t$ where $T(u)$ is an unknown $3\times 3$ matrix depending on $u$. Imposing the Einstein-Weyl equations (\ref{EW}) and using  (\ref{T1}) to eliminate the second order derivative $u_{tt}$ (no higher order derivatives of $u$ will occur in this calculation), we obtain a set of relations which have to vanish identically in the remaining partial derivatives of $u$. Thus, equating to zero terms at the  second order derivatives of $u$, we obtain $T$ in terms of $V$,
$$
T=2V^{-1}V'-(\ln \det V)'E,
$$
where $E$ is the $3\times 3$ identity matrix. This is equivalent to the formula (\ref{omega}) for $\omega$. The remaining terms vanish identically if and only if $V$ satisfies the constraint (\ref{b}). 
This finishes the proof of Theorem \ref{Thm3}.

\bigskip

\noindent{\bf Remark 1.} It was assumed in the proof of Theorem 3  that $\omega$ depends linearly on the first order derivatives of $u$. One can show that this assumption is unnecessary: even a rather general requirement that $\omega$ depends on some finite order jets of $u$ is already sufficiently restrictive, and leads to the formula (\ref{omega}) for $\omega$. 

\bigskip

\noindent{\bf Remark 2.} 
Equations from Theorem \ref{Thm1} possess implicit solutions  $u(x, y, t)$ of the form
 \begin{equation}
(x, y, t) M(u) (x, y, t)^T=1,
\label{central}
 \end{equation}
where $M(u)$ is a $3\times 3$ symmetric matrix of $u$. The level surfaces of such solutions, $u=const$, are central quadrics in the space of independent variables $x, y, t$. This construction is known as the central quadric ansatz \cite{Tod, Dunajski}. The  equation for  $M(u)$ is
$$
M'=sMVM/\sqrt{det M}, 
$$
$s=const$. It was demonstrated  in \cite{Tod, Dunajski} that in the cases of BF and dKP,  this equation  reduces to Painlev\'e trans\-cendents P3-P1. It was shown in \cite{Fer5} that  other integrable models from Theorem \ref{Thm1} lead to the remaining Painlev\'e equations P6-P4, with the full P6 corresponding to the first equation.  Thus, we obtain a whole  variety of Einstein-Weyl structures parametrized by Painlev\'e transcendents.

\section{Type II: quasilinear wave equations}

In this section we discuss geometric aspects of  quasilinear wave equations  (\ref{T2}),
$$
f_{11} u_{xx} + f_{22} u_{yy} + f_{33} u_{tt} + 2 f_{12} u_{xy} + 2f_{13} u_{xt} + 2f_{23} u_{yt} =0,
$$
here the coefficients $f_{ij}$ are assumed to be functions of the first order derivatives $u_x, u_y, u_t$ only. PDEs of this type were  investigated in \cite{Bur} based on their correspondence with conformal structures in projective space. It was pointed out that the moduli space of integrable equations  is 20-dimensional. In was shown   in \cite{Odesskii} that coefficients of the `generic' integrable equations of the form  (\ref{T2}) can be parametrized by generalized hypergeometric functions. We recall that the class of quasilinear wave equations  is invariant under the  group $GL(4)$ of linear  transformations of the variables $x^i,u$, 
where $(x^1, x^2, x^3)=(x, y, t)$. These transformations constitute the natural equivalence group of the problem.

The linearized equation has the form
$$
f_{ij}v_{x^ix^j}+\dots=0,
$$
and defines the conformal structure $g=g_{ij}dx^idx^j$ where  the matrix of $g_{ij}$ is the inverse of $f_{ij}$. Here dots denote terms with lower order derivatives of $v$. Our first  result is as follows.

 \begin{theorem}\label{Thm4} 
Equation (\ref{T2}) is linearizable by a transformation from the equivalence group  $GL(4)$ if and only if the conformal structure $g$ is conformally flat on every solution.
 \end{theorem}

\centerline{\bf Proof:}

\medskip

Let us first recall,  following \cite{Bur}, the linearizability conditions for equations of the form (\ref{T2}). Since the  structure of these conditions is the same  in any dimension, we will consider the general $n$-dimensional case,
$$
f_{ij}u_{x^ix^j}=0,
$$
where $f_{ij}$ are functions of the first order derivatives $u_{x^k}$ only,  $i, j, k=1, \dots, n$. Setting $p_k=u_{x^k}$ we will write down a system of differential constraints for $f_{ij}({\bf p})$ which are necessary and sufficient for the linearizability of the equation under study by a transformation from the equivalence group $GL(n+1)$.
Let us introduce the object
 $$
a_{ijk}=\partial_{p_k}f_{ij}-(c_k+2s_k)f_{ij}-s_if_{kj}-s_jf_{ki},
$$
where
$$
\begin{array}{c}
s_k=\frac{f^{ij}}{(n+2)(1-n)}\left(\partial_{p_k}f_{ij}-n\partial _{p_j}f_{ik}\right), \\
\ \\
c_k=\frac{f^{ij}}{(n+2)(n-1)}\left((n+3)\partial_{p_k}f_{ij}-2(n+1)\partial _{p_j}f_{ik}\right).
\end{array}
$$
Then the linearizability is equivalent to the following two conditions:

\begin{itemize}

\item $a_{ijk}=0,$

\item $\partial_{p_j}s_i-s_is_j=0.$
\end{itemize}
Geometrically, these conditions are equivalent to the existence of a flat connection $\nabla$ (in {\bf p}-coordinates) with Christoffel symbols $\Gamma^i_{jk}=s_j\delta^i_k+s_k\delta^i_j$ such that
$
\nabla _kf_{ij}=c_kf_{ij},
$
see \cite{Bur} for more details.

Let us now require that the conformal structure $g=g_{ij}dx^idx^j$ is conformally flat on every solution. Calculating the Cotton tensor (\ref{cot}) (now we set  $n=3$), and using  (\ref{T2}) and its differential consequences to eliminate all higher order partial derivatives of $u$ which contain differentiation by $t$ more than once,  we obtain a complicated expression which has to vanish identically
in the remaining higher order derivatives of $u$. In particular, requiring that  coefficients at the remaining fourth order derivatives  vanish identically (no higher order derivatives of $u$ will occur in this calculation), we obtain the first linearizability condition, $a_{ijk}=0$. 

There are two ways to proceed: collecting terms at the lower order derivatives of $u$ one can obtain the second set of linearizability conditions (this, however, leads to quite complicated calculations). Another way is to point out that the condition $a_{ijk}=0$ alone is already sufficiently restrictive \cite{Bur, Safaryan, Akivis}, and implies that the PDE in question is either linearizable, or reducible to the equation for minimal hypersurfaces in a (pseudo) Euclidean space,
$$
[(\nabla u)^2-1]\triangle u-(\nabla u) H (\nabla u)^t=0,
$$
where $\nabla u=(u_{x^1}, ..., u_{x^n})$ is the gradient of $u$,  $\triangle$ is the Laplacian, and $H$ is the Hessian matrix of $u$. In the three-dimensional case we arrive at the equation
$$
\begin{array}{c}
(u_y^2+u_t^2-1)u_{xx}+(u_x^2+u_t^2-1)u_{yy}+(u_x^2+u_y^2-1)u_{tt}\\
\ \\
-2(u_xu_yu_{xy}+u_xu_tu_{xt}+u_yu_tu_{yt})=0.
\end{array}
$$
To complete the proof it remains to point out  that the corresponding conformal structure $g$ is not conformally flat on generic solutions, furthermore,  the equation itself is not linearizable (in fact, not even integrable for $n\geq 3$). This finishes the proof of Theorem \ref{Thm4}.

\medskip

\noindent{\bf Remark.} 
This proof, and the proof of Theorem \ref{Thm2}, generalize to any dimension $n>3$, 
with the only change that one needs the Weyl tensor of conformal curvature instead 
of the Cotton tensor.
For non-linearizable differential equations, the requirement of conformal flatness  
singles out a subclass of exact solutions of a given PDE, see \cite{Lychagin1}.

\bigskip

Our next goal is to prove that conformal structures corresponding to formal linearizations of {\it integrable}  equations of the form (\ref{T2}) give rise to the   Einstein-Weyl geometry. Let us begin with examples of known integrable PDEs.

\bigskip

\noindent {\bf Example 1.} The equation $u_xu_{yt}+u_yu_{xt}+u_tu_{xy}= 0$
constitutes the Euler-Lagrange equation for the Lagrangian density $u_xu_yu_t$ which was obtained in \cite{FKT} in the classification of first order integrable Lagrangians.

\noindent Conformal structure: $g=(u_xdx+u_ydy+u_tdt)^2-2u_x^2dx^2-2u_y^2dy^2-2u_t^2dt^2$.

\noindent Covector: $\omega=-4\frac{u_xu_{yt}}{u_yu_t}\,dx-4\frac{u_yu_{tx}}{u_tu_x}\,dy-4\frac{u_tu_{xy}}{u_xu_y}\,dt$.

\bigskip

\noindent {\bf Example 2.} The equation $\left( u_y p(u_t) \right)_x + \left( u_x p(u_t) \right)_y + \left( u_x u_y p'(u_t) \right)_t = 0$
constitutes the Euler-Lagrange equation for the Lagrangian density $u_xu_yp(u_t)$, which can be viewed as a deformation of Example 1. In this case the integrability conditions
reduce to a single fourth order ODE for $p$,
\begin{equation*}
    p''''(p^2{p''}-2pp'^2) - p^2{p'''}^2  + 2p{p'}{p''}{p'''}
          + 8{p'}^3{p'''} - 9{p'}^2{p''}^2 = 0.
\end{equation*}
It was shown in \cite{FO} that the general  solution to this ODE  is  a  modular form of weight one and level three  known as the Eisenstein series $E_{1, 3}$.

\noindent Conformal structure:
$$
g=(p'u_xdx+p'u_ydy+pdt)^2-2p'^2u_x^2dx^2-2p'^2u_y^2dy^2-2p^2dt^2-2pp''u_xu_ydxdy.
$$

\noindent Covector:
$$
\begin{array}{c}
\omega=2\Bigl(\frac{pp''{}^2-pp'p'''+p'{}^{2}p''}{p(pp''-2p'{}^2)}\,u_xu_{tt}
-2\frac{p'u_x}{p\,u_y}\,u_{ty}\Bigr)\,dx
 +2\Bigl(\frac{pp''{}^2-pp'p'''+p'{}^{2}p''}{p(pp''-2p'{}^2)}\,u_yu_{tt}
-2\frac{p'u_y}{p\,u_x}\,u_{tx}\Bigr)\,dy\\
+2 \Bigl(\frac{pp'''-3p'p''}{pp''-2p'{}^2}\,u_{tt}
+2\bigl(\frac{u_{tx}}{u_x}+\frac{u_{ty}}{u_y}\bigr)\Bigr)\,dt.
\end{array}
$$
This structure is Einstein-Weyl if and only if $p$ satisfies the above fourth order ODE.

\bigskip





The second main result of this Section is as follows:

 \begin{theorem}\label{Thm5}
Equation (\ref{T2}) is integrable by the method of hydrodynamic reductions if and only if the corresponding  conformal structure $g$ is Einstein-Weyl on every solution, with the covector $\omega$ given by (\ref{omega}).
 \end{theorem}

\centerline{\bf Proof:}

\medskip

We will give two proofs of this result. The first one is computational, based on the explicit calculation of the Einstein-Weyl constraints, and the integrability conditions as derived in \cite{Bur}. The second proof utilises the fact that any integrable PDE of the form (\ref{T2})  possesses a dispersionless Lax pair \cite{Bur}. We demonstrate that the existence of a Lax pair implies that the Weyl connection $\mathbb{D}$, specified by the conformal structure $g=g_{ij}dx^idx^j$ and the covector (\ref{omega}), possesses a two-parameter family of null totally geodesic surfaces, the property known to be characteristic of  the Einstein-Weyl geometry \cite{Cartan}.

The first proof can be summarized as follows. Given an equation of the form (\ref{T2}), the conformal structure of its formal linearization is  $g=g_{ij}dx^idx^j$, where $g_{ij}$ is the inverse of $f_{ij}$. 
We will seek a covector $\omega=\omega_kdx^k$ in the form
$\omega_k=T^{ij}_ku_{x^ix^j}$, where $T^{ij}_k$ are certain functions of the first order derivatives of $u$. Imposing the Einstein-Weyl equations (\ref{EW}) and using  (\ref{T2}) and its differential consequences to eliminate all higher order derivatives of $u$ that contain differentiation by $t$ more than once (maximum third order derivatives of $u$ will occur in this calculation), we obtain a set of relations which have to vanish identically in the remaining partial derivatives of $u$. Thus, equating to zero terms at the  third order derivatives of $u$, we obtain the expression (\ref{omega}) for $\omega$. The remaining terms vanish identically if and only if the coefficients $f_{ij}$ satisfy the set of integrability conditions as derived in \cite{Bur}. This finishes the first proof of Theorem \ref{Thm5}.

\medskip

Let us now give a somewhat more conceptual (as well as less computational)  demonstration that the integrability is equivalent to the  Einstein-Weyl property. It is based on the fact that any integrable equation of the form (\ref{T2}) possesses a dispersionless Lax pair of the form
\begin{equation}
S_t=f(S_x, \, u_x, \, u_y, \, u_t), ~~~
S_y=g(S_x, \, u_x, \, u_y, \, u_t).
\label{Lax}
\end{equation}
This means that the consistency condition, $S_{ty}=S_{yt}$, is equivalent to the equation  (\ref{T2}). 
Lax pairs of this form are known to arise  as dispersionless limits of solitonic Lax pairs in $2+1$ dimensions \cite{Zakharov}.
Let us first outline the general construction which leads from the Lax pair (\ref{Lax}) to totally geodesic null surfaces of the Weyl connection $\mathbb{D}$. Differentiating (\ref{Lax}) by $x$ and setting $S_x=\lambda,\ u_x=a,\ u_y=b, \ u_t=c$ we obtain
\begin{equation}
\lambda_t=f_{\lambda} \lambda_x+f_{a}a_x+f_{b}b_x+f_cc_x, ~~~
\lambda_y=g_{\lambda} \lambda_x+g_{a}a_x+g_{b}b_x+g_cc_x.
\label{lambda}
\end{equation}
With this system we associate the  vector fields
$$
X=\frac{\partial}{\partial t}-f_{\lambda}\frac{\partial}{\partial x}+(f_{a}a_x+f_{b}b_x+f_cc_x)\frac{\partial}{\partial \lambda}, ~~~
Y=\frac{\partial}{\partial y}-g_{\lambda}\frac{\partial}{\partial x}+(g_{a}a_x+g_{b}b_x+g_cc_x)\frac{\partial}{\partial \lambda},
$$
which live in the extended four-dimensional space with coordinates $x, y, t, \lambda$. Note that the compatibility condition $\lambda_{ty}=\lambda_{yt}$ is equivalent to the commutativity of these vector fields: $[X, Y]=0$. 
The geometry behind this construction is as follows. Let us consider the cotangent bundle $Z$ of the solution $u(x, y, t)$, with local coordinates
$(x, y, t, S_x, S_y, S_t)$. Equations (\ref{Lax}) specify a four-dimensional submanifold $M^4\subset Z$ parametrised by $x, y, t$ and $ \lambda$. The compatibility of the equations (\ref{Lax}) indicates that this submanifold is coisotropic. The vector fields $X, Y$  generate  the kernel of the restriction to $M^4$ of the symplectic form $dS_x\wedge dx+dS_y\wedge dy+dS_t\wedge dt$. Equations (\ref{lambda}) mean that the vectors $X, Y$ are tangential to the hypersurface of $M^4$ defined by the equation $\lambda=\lambda(x, y, t)$.  

Projecting the two-parameter family of integral surfaces of the distribution spanned by $X, Y$  to the space of independent variables $x, y, t$ we obtain a two-parameter family of  null totally geodesic  surfaces of the Weyl connection $\mathbb{D}$. To see this we first project $X$ and $Y$. This  gives two vector fields
$$
\hat X=\frac{\partial}{\partial t}-f_{\lambda}\frac{\partial}{\partial x}, ~~~
\hat Y=\frac{\partial}{\partial y}-g_{\lambda}\frac{\partial}{\partial x},
$$
which commute if and only if $\lambda$ satisfies the equations (\ref{lambda}). It remains to show  that $\hat X$ and $\hat Y$  form a null distribution (that is, tangential to the null cones of $g$), and  that the covariant derivatives  $\mathbb{D}_{\hat X}\hat X, \ \mathbb{D}_{\hat X}\hat Y,\ \mathbb{D}_{\hat Y} \hat X, \ \mathbb{D}_{\hat Y}\hat Y $ belong to the span of $\hat X, \hat Y$. 
Equivalently, one can introduce the covector $\theta=dx+g_{\lambda}dy+ f_{\lambda}dt$ which annihilates $\hat X, \hat Y$, and verify that  $\theta$ is null, and that $\mathbb{D}_{\hat X}\theta \wedge \theta=\mathbb{D}_{\hat Y}\theta \wedge \theta=0$.
This follows from the equations satisfied by the functions $f(\lambda, a, b, c)$ and $g(\lambda, a, b, c)$ as derived in \cite{Bur}:
\begin{align}\label{FG}
     f_a &= 2kf_{12}+g_{\lambda}kf_{22}+f_{\lambda}(kf_{23}-p), ~~~ &g_a& =-2kf_{13}-f_{\lambda}kf_{33}-g_{\lambda}(kf_{23}+p) , \notag\\
     f_b &= kf_{22}, ~~~ &g_b& = -kf_{23}+p,\\
     f_c &= kf_{23}+p, ~~~ &g_c &= -kf_{33}, \notag
\end{align}
where $p(\lambda, a, b, c)$  and $k(\lambda, a, b, c)$ are yet another two auxiliary functions. Furthermore, $f_{\lambda}$ and $g_{\lambda}$  satisfy the  relation
\begin{equation}
f_{11} + f_{22}g_{\lambda}^2 + f_{33}f_{\lambda}^2 + 2f_{12}g_{\lambda} + 2f_{13}f_{\lambda}
 + 2f_{23}f_{\lambda}g_{\lambda}=0,
\label{disp1}
\end{equation}
which means that the covector $\theta$ is null. To close the system (\ref{FG}) -- (\ref{disp1}) one proceeds as follows. Calculating the consistency conditions for Eqs. (\ref{FG}), $f_{ab}=f_{ba}$,  $g_{ab}=g_{ba}$,  etc, six conditions altogether, and differentiating the  relation (\ref{disp1}) by $a, b, c$ and $\lambda$, one obtains ten relations which can be solved for $f_{\lambda \lambda}, g_{\lambda \lambda}$ and the first order derivatives of $k$ and $p$.  Modulo these relations, the conditions $\mathbb{D}_{\hat X}\theta \wedge \theta=\mathbb{D}_{\hat Y}\theta \wedge \theta=0$ are satisfied identically. 
This finishes the second proof of Theorem \ref{Thm5}.

\section{Type III:  dispersionless Hirota equations}

In this section we discuss geometric aspects of  PDEs  (\ref{T3}) of the dispersionless Hirota type,
$$
F(u_{xx}, u_{xy}, u_{yy}, u_{xt}, u_{yt}, u_{tt})=0,
$$
which were  investigated in \cite{MaksEgor, Fer4}, revealing a remarkable correspondence with hypersurfaces of the Lagrangian Grassmanian. Geometric aspects of $GL(2)$ structures associated with such equations were studied in \cite{Smith}. In was shown   in \cite{Odesskii} that the `generic' integrable equation of the form  (\ref{T3}) can be parametrized by generalized hypergeometric functions. Recall that equations of the form (\ref{T3}) are invariant under the group $Sp(6)$ of linear symplectic transformations of the variables $x^i,u_{x^i}$, where $(x^1, x^2, x^3)=(x, y, t)$. These transformations constitute the natural equivalence group of the problem.

The linearized equation has the form
$$
F_{{ij}}v_{x^ix^j}=0,
$$
and  defines a conformal structure $g=g_{ij}dx^idx^j$ where the matrix of $g_{ij}$ is the inverse of $F_{ij}=\partial F/\partial u_{x^ix^j}$. Our first result is as follows.

 \begin{theorem}\label{Thm6} 
Equation (\ref{T3}) is linearizable by a transformation from the equivalence group $Sp(6)$ if and only if the conformal structure $g$ is conformally flat on every solution.
 \end{theorem}

\centerline{\bf Proof:}

\medskip

\noindent Solving for $u_{tt}$ (we can always bring our equation into the form with a non-trivial dependence on $u_{tt}$  due to the presence of a large equivalence group), we can rewrite (\ref{T3}) in the form
\begin{equation}
u_{tt}=f(u_{xx}, u_{xy}, u_{yy}, u_{xt}, u_{yt}).
\label{T31}
\end{equation}
The corresponding linearized equation is
$$
v_{tt}=f_{u_{xx}}v_{xx}+f_{u_{xy}}v_{xy}+f_{u_{yy}}v_{yy}+f_{u_{xt}}v_{xt}+f_{u_{yt}}v_{yt},
$$
with the associated conformal structure (with upper indices) defined by the matrix
$$
P=\left(
\begin{array}{ccc}
f_{u_{xx}}&\frac{1}{2}f_{u_{xy}}&\frac{1}{2}f_{u_{xt}}\\
\ \\
\frac{1}{2} f_{u_{xy}}&f_{u_{yy}}&\frac{1}{2}f_{u_{yt}}\\
\ \\
\frac{1}{2}f_{u_{xt}}&\frac{1}{2}f_{u_{yt}}&-1
\end{array}
\right).
$$
We  require  this structure to be conformally flat for every background solution $u(x, y, t)$. Calculating the Cotton tensor (\ref{cot}) and using  (\ref{T31}) and its differential consequences to eliminate all higher order derivatives of $u$ containing differentiation by $t$ more than once,  we obtain a complicated expression which has to vanish identically
in the remaining higher order derivatives of $u$ (maximum fifth order derivatives  will appear in this calculation). In particular, requiring that  coefficients at the remaining fifth order derivatives of $u$ vanish identically, we obtain  nine second order differential constraints for $f$:
\begin{equation}
\label{constr}
\begin{array}{c}
f_{u_{xx}}f_{u_{xt}u_{xt}}+f_{u_{xx}u_{xx}}=0, ~~~ f_{u_{yy}}f_{u_{yt}u_{yt}}+f_{u_{yy}u_{yy}}=0,\\
\ \\
f_{u_{xt}}f_{u_{xt}u_{xt}}+2f_{u_{xt}u_{xx}}=0, ~~~ f_{u_{yt}}f_{u_{yt}u_{yt}}+2f_{u_{yt}u_{yy}}=0,\\
\ \\
f_{u_{yt}}f_{u_{xt}u_{xt}}+2(f_{u_{xt}}f_{u_{xt}u_{yt}}+f_{u_{xt}u_{xy}}+f_{u_{yt}u_{xx}})=0,\\
\ \\
f_{u_{xt}}f_{u_{yt}u_{yt}}+2(f_{u_{yt}}f_{u_{xt}u_{yt}}+f_{u_{yt}u_{xy}}+f_{u_{xt}u_{yy}})=0,\\
\ \\
f_{u_{xy}}f_{u_{xt}u_{xt}}+2f_{u_{xx}}f_{u_{xt}u_{yt}}+2f_{u_{xx}u_{xy}}=0,\\
\ \\
f_{u_{xy}}f_{u_{yt}u_{yt}}+2f_{u_{yy}}f_{u_{xt}u_{yt}}+2f_{u_{yy}u_{xy}}=0,\\
\ \\
f_{u_{yy}}f_{u_{xt}u_{xt}}+f_{u_{xx}}f_{u_{yt}u_{yt}}+2f_{u_{xy}}f_{u_{xt}u_{yt}}+2f_{u_{xx}u_{yy}}+f_{u_{xy}u_{xy}}=0.
\end{array}
\end{equation}
It was  shown  in \cite{Ruggeri, Boillat, Colin}  that  these relations characterize symplectic Monge-Amp\`ere equations, that is, PDEs  (\ref{T3}) such that the left hand side $F$  can be represented as a  linear combination of all possible minors of the Hessian matrix of $u$,
$$
U=\left(
\begin{array}{ccc}
u_{xx} & u_{xy} & u_{xt}\\
u_{xy} & u_{yy} & u_{yt}\\
u_{xt} & u_{yt} & u_{tt}\\
\end{array}
\right).
$$
Symplectic Monge-Amp\`ere equations and  differential constraints (\ref{constr}) have a clear geometric interpretation. Let us consider the Lagrangian Grassmannian $\Lambda^6$ which can be (locally) parametrised by $3\times 3$ symmetric matrices $U$. 
Minors of $U$ define the Pl\"ucker embedding of $\Lambda^6$ into  projective space $P^{13}$. We will identify $\Lambda^6$ with the image of this projective embedding. Symplectic Monge-Amp\`ere equations can be viewed as hyperplane sections $M^5$ of $\Lambda^6\subset P^{13}$. We point out that  differential constraints (\ref{constr}) can be represented in compact form as
\begin{equation}
\label{m31}
\begin{array}{c}
d^2f=2a_0(dfdu_{xy}-du_{xt}du_{yt})+2a_1(dfdu_{yy}-(du_{yt})^2)+2a_2(dfdu_{xx}-(du_{xt})^2)\\
\ \\
+2b_0(du_{xx}du_{yy}-(du_{xy})^2)+2b_1(du_{xt}du_{xy}-du_{yt}du_{xx})+2b_2(du_{yt}du_{xy}-du_{xt}du_{yy})
\end{array}
\end{equation}
indeed, they follow from (\ref{m31}) on elimination of the coefficients $a_i, b_i$. Here $d^2f$ is the symmetric differential of $f$. Notice that $d^2f$ and the six quadratic expressions on the right hand side of (\ref{m31}) are nothing but second fundamental forms of the submanifold $M^5\subset \Lambda^6\subset P^{13}$ defined by (\ref{T31}). Furthermore, the six fundamental forms on the right hand side of (\ref{m31}) are the restrictions to $M^5$ of the second fundamental forms of $\Lambda^6\subset P^{13}$. Thus, (\ref{m31}) says that $M^5$ has no nontrivial second fundamental forms `of its own', that is, all its second fundamental forms  can be obtained as restriction of the second fundamental forms of $\Lambda^6$. This is an obvious necessary condition for a submanifold $M^5\subset \Lambda^6\subset P^{13}$ to be a hyperplane section. In the present case, it is also sufficient. 
Calculating consistency conditions for  (\ref{m31})  we obtain the equation for $a_i$ and $ b_i$,
\begin{equation}
\begin{array}{c}
da_0=a_0\varphi-2sdu_{12}, ~~~ da_1=a_1\varphi+sdu_{11}, ~~~da_2=a_2\varphi+sdu_{22}, \\
\ \\
db_0=b_0\varphi+sdf, ~~~ db_1=b_1\varphi+{2} s du_{02}, ~~~db_2=b_2\varphi+{2}sdu_{01}, ~~~  ds=s\varphi.
\end{array}
\label{m51}
\end{equation}
 Here $\varphi=a_0du_{xy}+a_1du_{yy}+a_2du_{xx}$, and $s$ is yet another auxiliary function. One can verify that $d\varphi =0$. Equations (\ref{m31}) and (\ref{m51}) constitute an  involutive differential system for $f$  which characterizes symplectic Monge-Amp\`ere equations.

Once we know that our PDE  is of symplectic Monge-Amp\`ere type, there are two ways to proceed. The first one is to use the fact that linearizable Monge-Amp\`ere equations correspond to special hyperplane sections of $\Lambda^6$ such that the corresponding hyperplane is tangential to $\Lambda^6$, that is, belongs to the dual variety \cite{Fer4, DF}.  Written in differential form, this simple geometric property gives just one extra condition which can be used to express $s$   in terms of $a_i, b_i$. The resulting formula  is quite complicated, reflecting the fact the the dual variety of $\Lambda^6$ is a quartic hypersurface defined by a rather cumbersome equation. Explicitly, we have:
$$
\begin{array}{c}
 s=-\{a_0^2f_{u_{xx}}f_{u_{yy}}+a_1^2f_{u_{xx}}^2+a_2^2f_{u_{yy}}^2-a_0a_1f_{u_{xx}}f_{u_{xy}}-a_0a_2f_{u_{xy}}f_{u_{yy}}+a_1a_2(f_{u_{xy}}^2-2f_{u_{xx}}f_{u_{yy}})-\\
\ \\
a_0b_0(f_{u_{xt}}f_{u_{yt}}+f_{u_{xy}})+a_1b_1(f_{u_{xt}}f_{u_{xy}}-f_{u_{yt}}f_{u_{xx}})+a_2b_2(f_{u_{yt}}f_{u_{xy}}-f_{u_{xt}}f_{u_{yy}})+\\
\ \\
a_1b_0(f_{u_{xt}}^2+2f_{u_{xx}})+a_2b_0(f_{u_{yt}}^2+2f_{u_{yy}})-a_0b_1f_{u_{xt}}f_{u_{yy}}-a_0b_2f_{u_{yt}}f_{u_{xx}}+\\
\ \\
a_1b_2f_{u_{xt}}f_{u_{xx}}+a_2b_1f_{u_{yt}}f_{u_{yy}}+b_0^2-b_1^2f_{u_{yy}}-b_2^2f_{u_{xx}}-b_0b_1f_{u_{yt}}-b_0b_2f_{u_{xt}}-b_1b_2f_{u_{xy}}\}/4\det P,
\end{array}
$$
here the matrix $P$ was defined at the beginning of the proof.
A direct calculation shows that the remaining coefficients of the  Cotton tensor vanish if and only if the above linearizability condition holds. This, however, is a rather complicated calculation. 

Another way is to use the fact that, in three dimensions, any
non-degenerate symplectic Monge-Amp\`ere equation is either linearizable, or
$Sp(6)$-equivalent to one of the three canonical forms,
 \begin{equation}\label{MA}
\Hess u=1,\qquad  \Hess u=u_{xx}+u_{yy}+u_{tt}, \qquad  \Hess
u=u_{xx}+u_{yy}-u_{tt},
 \end{equation}
see  \cite{Lychagin, Banos}. Note that the first equation governs improper affine hyperspheres,
while the last two describe special Lagrangian 3-folds.
One can verify by a direct calculation that formal linearizations of the equations (\ref{MA})  are not conformally flat for generic solutions. Thus, once again conformal flatness proves to be equivalent to the linearizability.  This finishes the proof of Theorem~\ref{Thm6}.

\bigskip

As in the previous cases, formal linearizations of {\it integrable} equations of the form (\ref{T3}) give rise to conformal structures satisfying the  Einstein-Weyl property. 

\bigskip

\noindent {\bf Example 1.} The equation $e^{u_{tt}}=e^{u_{xx}}+e^{u_{yy}}$ appeared in \cite{Fer4}  in the classification of integrable PDEs of the form $F(u_{xx}, u_{yy}, u_{tt})=0$.

\noindent Conformal structure: $g=e^{-u_{xx}}dx^2+e^{-u_{yy}}dy^2-e^{-u_{tt}}dt^2$.

\noindent Covector: $\omega=2(u_{ttt}dt+u_{xxx}dx+u_{yyy}dy)-d(u_{tt}+u_{xx}+u_{yy})$.
\bigskip

\noindent {\bf Example 2.} The equation $u_{tt} =\frac{u_{xy}}{u_{xt}}+\frac{1}{6}\eta (u_{xx})u_{xt}^{2}$ appeared in  \cite{MaksEgor} in the classification of integrable hydrodynamic chains.
Here the integrability conditions reduce to  the Chazy equation for $\eta$,
$\eta^{\prime \prime \prime }+2\eta \eta ^{\prime \prime }=3(\eta^{\prime})^2$.

\noindent Conformal structure: $g=4u_{xt}dxdy-\left(\frac{2}{3}\eta'u_{xt}^4+s^2\right)dy^2+2sdydt-dt^2$, here $s=\frac{1}{3}\eta u_{xt}^2-\frac{u_{xy}}{u_{xt}}$.

\noindent Covector:
$$\begin{array}{c}
\omega=
\Bigl[(\tfrac23\,u_{{tx}}\eta+4u_{xy}u_{tx}^{-2})u_{ttx}
+(\tfrac29\,u_{tx}^2\eta^2+\tfrac83\,u_{tx}^2\eta' -u_{xy}^2u_{tx}^{-4}
-\tfrac13\,u_{xy}u_{tx}^{-1}\eta) u_{txx}\\
+ (\tfrac19\,u_{tx}^3\eta\,\eta'+\tfrac23\,u_{tx}^3\eta''-\tfrac13\,u_{xy}\eta') u_{xxx}
+(u_{xy}u_{tx}^{-3}-\tfrac13\,\eta) u_{xxy}
-2\,u_{tx}^{-1}u_{txy}
\Bigr]\,dy\\
-\left[(u_{xy}u_{tx}^{-3}+\tfrac23\eta)u_{txx}+\tfrac13\eta'\,u_{tx}u_{xxx}
-u_{t,x}^{-2}u_{xxy}-2u_{tx}^{-1}u_{ttx}\right]\,dt.
\end{array}
$$
This structure is Einstein-Weyl if and only if $\eta$ solves the Chazy equation.

The second main result of this Section is as follows:

 \begin{theorem}\label{Thm7}
Equation (\ref{T3}) is integrable by the method of hydrodynamic reductions if and only if the corresponding  conformal structure $g$ is Einstein-Weyl on every solution, with the covector $\omega$ given by (\ref{omega}).
 \end{theorem}

\centerline{\bf Proof:}

\medskip

Given an equation of the form (\ref{T3}), the conformal structure of its formal linearization is  $g=g_{ij}dx^idx^j$,
where $g_{ij}$ is the inverse of $F_{ij}$. We will seek a covector $\omega=\omega_kdx^k$ in the form
$\omega_k=T^{ijl}_ku_{x^ix^jx^l}$ where $T^{ijl}_k$ are certain functions of the second order derivatives of $u$. Imposing the Einstein-Weyl equations (\ref{EW}) and using  (\ref{T3}) and its differential consequences to eliminate all higher order derivatives of $u$ that contain differentiation by $t$ more than once (we use the representation (\ref{T31}); maximum fourth order derivatives of $u$ will occur in this calculation), we obtain a set of relations which have to vanish identically in the remaining partial derivatives of $u$. Thus, equating to zero coefficients at the fourth order derivatives of $u$, we obtain the expression (\ref{omega}) for $\omega$.

With this expression, $R-\Lambda\,g$ is purely quadratic in the third order derivatives $u_{x^ix^jx^l}$.
Choosing $\Lambda$ in such a way that $dt^2$ term disappears, we get
$5\cdot 28=140$ coefficients at these quadratic terms which are third order differential polynomials in $f$. Note that these coefficients are linear in the third order derivatives of $f$.
Their vanishing is equivalent to some 35 identities constituting 
an involutive closed system of third order PDEs for $f$. 
These are precisely the integrability conditions as derived in \cite{Fer4}. 
This finishes the proof of Theorem \ref{Thm7}.

\section{Type IV: systems of hydrodynamic type}

In this section we discuss geometric aspects of  integrable systems of hydrodynamic type  (\ref{T4}),
$$
A(u)u_x+B(u)u_y+C(u)u_t=0,
$$
where $u=(u^1, u^2)^t$ is a two-component column vector of the dependent variables, and $A(u), B(u), C(u)$ are $2\times 2$ matrices. It will be assumed that there is a matrix in the span of $A, B, C$ that is hyperbolic. This class is invariant under arbitrary changes of variables $u^1, u^2$, as well as linear transformations of $x, y, t$, which constitute the natural equivalence group of the problem. If $C$ is non-degenerate, the multiplication by $C^{-1}$ brings the system into evolutionary form.  The integrability of  systems of hydrodynamic type was investigated in \cite{FK}. In was shown   in \cite{Odesskii} that `generic' integrable system of the form  (\ref{T4}) can be parametrized by generalized hypergeometric functions. 
The linearized system has the form
$$
A(u)v_x+B(u)v_y+C(u)v_t+\dots=0,
$$
where dots denote terms which do not contain derivatives of $v$. The corresponding dispersion relation (which coincides with the equation for characteristic covectors) is given by the formula
$$
\det(\lambda^1A(u)+\lambda^2B(u)+\lambda^3C(u))=0.
$$
This is a quadratic form in $\lambda$ which can be represented as $(\lambda^1, \lambda^2, \lambda^3)D(u)(\lambda^1, \lambda^2, \lambda^3)^t$ where $D$ is a $3\times 3$ symmetric matrix. It defines the conformal structure $g=g_{ij}dx^idx^j$ where the matrix of $g_{ij}$ is the inverse of $D$. Recall that $(x^1, x^2, x^3)=(x, y, t)$. Our first result is as follows:

 \begin{theorem}\label{Thm8} 
System (\ref{T4}) is linearizable by a transformation from the equivalence group  if and only if it is integrable, and the conformal structure $g$ is conformally flat on every solution (note that conformal flatness alone is no longer sufficient for the linearizability).
 \end{theorem}

\centerline{\bf Proof:}

\medskip

\noindent Without any loss of generality one can assume that system (\ref{T4}) is represented in the form 
\begin{equation}
\left(
\begin{array}{c}
u^1 \\
\ \\
u^2
\end{array}
\right)_t+
\left(
\begin{array}{cc}
a & 0 \\
\ \\
0 & b
\end{array}
\right)
\left(
\begin{array}{c}
u^1 \\
\ \\
u^2
\end{array}
\right)_x+
\left(
\begin{array}{cc}
p&q \\
\ \\
r&s
\end{array}
\right)
\left(
\begin{array}{c}
u^1 \\
\ \\
u^2
\end{array}
\right)_y=0
\label{sys}
\end{equation}
 (multiply by $C^{-1}$  and use a change of variables $u^1, u^2$ to  make the matrix $C^{-1}A$
diagonal: such diagonalization is  always possible in the two-component hyperbolic
situation). 
Here the matrix elements $a, b, p, q, r, s$ are functions of $u^1, u^2$. The corresponding dispersion relation takes the form
\begin{equation}
(\lambda^1a+\lambda^2p+\lambda^3)(\lambda^1b+\lambda^2s+\lambda^3)-qr(\lambda^2)^2=0,
\label{dr}
\end{equation}
with the associated  matrix
$$
D(u)=\left(
\begin{array}{ccc}
ab&\frac{as+bp}{2}&\frac{a+b}{2}\\
\ \\
\frac{as+bp}{2} &ps-qr&\frac{p+s}{2}\\
\ \\
\frac{a+b}{2}&\frac{p+s}{2}&1
\end{array}
\right).
$$
This gives rise to the conformal structure $g=g_{ij}dx^idx^j$ where the matrix of $g_{ij}$ is the inverse of $D$. We  require $g$ to be conformally flat for every background solution $u$. Calculating the Cotton tensor (\ref{cot}) and using  (\ref{sys}) and its differential consequences to eliminate all higher order derivatives of $u$ which contain differentiation by $t$,  we obtain the expression which has to vanish identically
in the remaining higher order derivatives of $u$ (maximum third order derivatives  will appear in this calculation). In particular, requiring that  coefficients at the remaining third order derivatives of $u$ vanish identically, we obtain that $a, b, p, s$ and $qr$ must be constant (thus, all coefficients of the dispersion relation are constants). In this case the rest of the Cotton tensor vanishes identically. By a transformation from the equivalence group any such system can be brought to the form
$$
u^1_t=qu^2_y, ~~~ u^1_y={q}u^2_x,
$$
where $q$ is still an arbitrary function of $u^1, u^2$. For generic $q$ such systems are neither linearizable nor integrable. Imposing the integrability conditions as derived in \cite{FK} we obtain just one extra constraint, $(\ln q)_{u^1u^2}=0,$
which is equivalent to the existence of a change of variables $u^1\to \varphi^1(u^1), \ u^2\to \varphi^2(u^2)$ bringing system (\ref{sys}) to a constant coefficient form. This finishes the proof of Theorem~\ref{Thm8}.

\bigskip

Formal linearizations of {\it integrable} systems of the form (\ref{T4}) give rise to conformal structures satisfying the  Einstein-Weyl property. 

\medskip

\noindent {\bf Example 1.} Let us consider the system
$$
u^1_t+\frac{u^1u^1_x+u^2u^2_x}{(u^1)^2+(u^2)^2}+\frac{u^1u^2_y-u^2u^1_y}{(u^1)^2+(u^2)^2}=0, ~~~
u^2_t+\frac{u^2u^1_x-u^1u^2_x}{(u^1)^2+(u^2)^2}+\frac{u^1u^1_y+u^2u^2_y}{(u^1)^2+(u^2)^2}=0,
$$
which was obtained in \cite{Fer1} as the first order form of the Boyer-Finley equation: indeed,  the expression 
$\rho=(u^1)^2+(u^2)^2$ satisfies  the equation
$$
 \rho_{tt}=\triangle \ln \rho,
$$
$\triangle=\partial_x^2+\partial_y^2$. In this case the dispersion relation is
$$
\frac{(\lambda^1)^2+(\lambda^2)^2}{\rho}-(\lambda^3)^2=0.
$$
The corresponding Einstein-Weyl structure takes the familiar form \cite{Ward}:

\noindent Conformal structure: $g=\rho(dx^2+dy^2)-dt^2$.

\noindent Covector: $\omega=2\frac{\rho_t}{\rho}dt$.
\bigskip

\medskip

\noindent {\bf Example 2.} A class of Hamiltonian systems of hydrodynamic type can be represented  in the form
\begin{equation}
v_t+(H_v)_y=0, ~~~ w_x+(H_w)_y=0,
\label{vw}
\end{equation}
here  $H(v, w)$ is the Legendre transform of the Hamiltonian density.  The  dispersion relation,
$$
(\lambda^2H_{vv}+\lambda^3)(\lambda^2H_{ww}+\lambda^1)-(\lambda^2H_{vw})^2=0,
$$
gives rise to the conformal structure
$$
g=(dy-H_{ww}dx-H_{vv}dt)^2-4H_{vw}^2dxdt.
$$
The corresponding covector $\omega$ is given by the formula (\ref{omega}),
 \begin{gather*}
H_{vw}^2 \omega=\Bigl(
H_{ww}\Delta_y+2H_{vw}^2(H_{ww})_y+H_{ww}\bigl((H_{vv})_x+(H_{ww})_t\bigr)
\Bigr) dx\\
-\Bigl(\Delta_y+(H_{vv})_x+(H_{ww})_t\Bigr) dy+
\Bigl(
H_{vv}\Delta_y+2H_{vw}^2(H_{vv})_y+H_{vv}\bigl((H_{vv})_x+(H_{ww})_t\bigr)  
\Bigr) dt,
 \end{gather*}
here $\Delta=H_{vv}H_{ww}-2H_{vw}^2$.
One can verify that $g, \omega$ satisfy the Einstein-Weyl constraints if and only if the potential $H(v, w)$  satisfies the set of integrability conditions as derived in \cite{Fer1} based on the method of hydrodynamic reductions:
\begin{equation}
\begin{array}{c}
H_{vw}H_{vvvv}=2H_{vvv}H_{vvw}, \\
\ \\
H_{vw}H_{vvvw}=2H_{vvv}H_{vww}, \\
\ \\
H_{vw}H_{vvww}=H_{vvw}H_{vww}+H_{vvv}H_{www}, \\
\ \\
H_{vw}H_{vwww}=2H_{vvw}H_{www}, \\
\ \\
H_{vw}H_{wwww}=2H_{vww}H_{www}.
\end{array}
\label{H}
\end{equation}
This system is in involution. It was shown in \cite{FMS} that its `generic' solution  is given by the formula
\begin{equation}
H(v, w)=Z(v+w)+\epsilon Z(v+\epsilon w)+\epsilon^2 Z(v+\epsilon^2
w);
 \label{gen}
\end{equation}
here $\epsilon=e^{2\pi i/3}$, and $Z''(s)=\zeta(s)$ where $\zeta $ is the
Weierstrass zeta-function: $\zeta'=-\wp, \ (\wp')^2=4\wp^3-g_3$ (equianharmonic case $g_2=0$). Degenerations of this solution  correspond to
$$
H(v, w)=\frac{1}{2}v^2\zeta (w), ~~~
H(v, w)=(v+w)\ln (v+w), 
$$
as well as the following  polynomial potentials:
$$
H(v, w)=v^2w^2,
~~~ H(v, w)=vw^2+\frac{\alpha}{5}w^5, ~~~ 
H(v, w)=vw+\frac{1}{6}w^3.
$$

\medskip

The second main result of this Section is as follows:

 \begin{theorem}\label{Thm9}
System (\ref{T4}) with  nonconstant dispersion relation is integrable by the method of hydrodynamic reductions if and only if the corresponding  conformal structure $g$ is Einstein-Weyl on every solution, with the covector $\omega$ given by (\ref{omega}).
 \end{theorem}

\centerline{\bf Proof:}

\medskip

We will give two proofs of this result. The first one is computational, based on the explicit calculation of the Einstein-Weyl constraints, and the integrability conditions as derived in \cite{FK}. The second proof utilises the fact that any two-component integrable system of hydrodynamic type possesses a dispersionless Lax pair \cite{FK}. 


The first proof can be summarized as follows. We  consider system (\ref{T4}) represented in the form (\ref{sys}). The associated conformal structure is $g=g_{ij}dx^idx^j$
where $g_{ij}$ is the inverse of $D$. We will seek a covector $\omega=\omega_kdx^k$ in the form
$\omega_k=T^i_{kj}u^j_{x^i}$ where $T^{i}_{kj}$ are certain functions of  $u^1, u^2$. Imposing the Einstein-Weyl equations (\ref{EW}) and using  (\ref{sys}) and its differential consequences to eliminate all higher order derivatives of $u$ that contain differentiation by $t$ (maximum second order derivatives of $u$ will occur in this calculation), we obtain a set of relations which have to vanish identically in the remaining partial derivatives of $u$. Thus, equating to zero coefficients at the second order derivatives of $u$, we obtain the expression (\ref{omega}) for $\omega$. Equating to zero the remaining terms we obtain 15  relations providing all second order partial derivatives (in $u^1, u^2$) of the coefficients of the dispersion relation, that is,  of $a, b, p,  s$ and $qr$. These relations constitute part of the total set of  16 integrability conditions as derived in \cite{FK}. To recover the missing condition we 
calculate the compatibility conditions for the 15 relations at hand.  This leads to the two cases:

\noindent {\bf Case 1}:  all coefficients of the dispersion relation are constant, in this case all of the 15 relations are satisfied identically. Any such system can be brought to the form
$u^1_t=qu^2_y, ~ u^1_y={q}u^2_x,$
see the proof of Theorem \ref{Thm8}. 
Although the Einstein-Weyl property is trivially satisfied,   the system does not  need to be integrable  for generic $q(u^1, u^2)$. This is why we need to eliminate the case of constant dispersion relation from the statement of Theorem \ref{Thm9}.

\noindent {\bf Case 2}: one of the 15 relations, namely,  the expression for the mixed derivative $(qr)_{u^1u^2}$, splits into two separate expressions for $q_{u^1u^2}$ and $r_{u^1u^2}$. 
This gives all of the 16  integrability conditions as derived in \cite{FK} based on the method of hydrodynamic reductions, thus finishing the first proof. 

\medskip

As in Theorem \ref{Thm5}, there exists another  demonstration that the integrability is equivalent to the Einstein-Weyl property. It is based on the fact that any integrable system of hydrodynamic type, which we again assume represented in the form (\ref{sys}), possesses a dispersionless Lax pair 
\begin{equation}
S_t=f(S_y, \ u^1, \ u^2), ~~~ S_x=g(S_y, \ u^1, \ u^2).
\label{Lax1}
\end{equation}
This means that the consistency condition $S_{tx}=S_{xt}$ is equivalent to the system  (\ref{sys}). 
Differentiating (\ref{Lax1}) by $y$ and setting $S_y=\lambda$ we obtain
\begin{equation}
\lambda_t=f_{\lambda} \lambda_y+f_{u^1}u^1_y+f_{u^2}u^2_y, ~~~
\lambda_x=g_{\lambda} \lambda_y+g_{u^1}u^1_y+g_{u^2}u^2_y.
\label{lambda1}
\end{equation}
With this system we associate the pair of vector fields
$$
X=\frac{\partial}{\partial t}-f_{\lambda}\frac{\partial}{\partial y}+(f_{u^1}u^1_y+f_{u^2}u^2_y)\frac{\partial}{\partial \lambda}, ~~~
Y=\frac{\partial}{\partial x}-g_{\lambda}\frac{\partial}{\partial y}+(g_{u^1}u^1_y+g_{u^2}u^2_y)\frac{\partial}{\partial \lambda},
$$
which live in the extended four-dimensional space with coordinates $x, y, t, \lambda$. Note that the compatibility condition $\lambda_{tx}=\lambda_{xt}$ is equivalent to the commutativity of these vector fields. Projecting the two-parameter family of integral surfaces of the distribution spanned by $X, Y$ to the space of independent variables $x, y, t$ we obtain a two-parameter family of  null totally geodesic  surfaces of the Weyl connection $\mathbb{D}$. This can be seen as follows.
Equations (\ref{lambda1}) mean that the vectors $X, Y$ are tangential to the hypersurface $\lambda=\lambda(x, y, t)$.  
Projecting $X$ and $Y$ to the space of independent variables $x, y, t$ we obtain two vector fields
$$
\hat X=\frac{\partial}{\partial t}-f_{\lambda}\frac{\partial}{\partial y}, ~~~
\hat Y=\frac{\partial}{\partial x}-g_{\lambda}\frac{\partial}{\partial y},
$$
which commute if and only if $\lambda$ satisfies the equations (\ref{lambda1}). It remains to show  that $\hat X$ and $\hat Y$  form a null distribution (that is,  tangential to the null cones of the conformal structure $g$), and  that the covariant derivatives  $\mathbb{D}_{\hat X}\hat X, \ \mathbb{D}_{\hat X}\hat Y,\ \mathbb{D}_{\hat Y} \hat X, \ \mathbb{D}_{\hat Y}\hat Y $ belong to the span of $\hat X, \hat Y$. 
Equivalently, one can introduce the covector $\theta=g_{\lambda}dx+dy+ f_{\lambda}dt$ which annihilates $\hat X, \hat Y$, and verify  that  $\theta$ is null, and that $\mathbb{D}_{\hat X}\theta \wedge \theta=\mathbb{D}_{\hat Y}\theta \wedge \theta=0$.
This   follows from the equations satisfied by the functions $f(\lambda, u^1, u^2)$ and $g(\lambda, u^1, u^2)$ as derived in \cite{FK}:
$$
\begin{array}{c}
f_{u^1}=-a\ g_{u^1}, ~~~ f_{u^2}=-b\ g_{u^2}, \\
\ \\
 f_{\lambda}=\frac{{b\left(p+r\frac{g_{u^2}}{g_{u^1}}\right)-a\left(s+q\frac{g_{u^1}}{g_{u^2}}\right)}}{{a-b}},
~~~
  g_{\lambda}=\frac{{s+q\frac{g_{u^1}}{g_{u^2}}-p-r\frac{g_{u^2}}{g_{u^1}}}}{{a-b}}, \\
  \ \\
g_{u^1u^2}=\frac{a_{u^2}}{b-a}\ g_{u^1}+\frac{b_{u^1}}{a-b}\ g_{u^2}, \\
\ \\
g_{u^1u^1}=\frac{g_{u^1}[g_{u^2}^2(r(b_{u^1}-a_{u^1})+(a-b)r_{u^1})+g_{u^1}g_{u^2}((a-b)p_{u^1}+(s-p)a_{u^1}-ra_{u^2})+qa_{u^1}g_{u^1}^2]}{(a-b) 
r g_{u^2}^2}, \\
\  \\
g_{u^2u^2}=\frac{g_{u^2}[g_{u^1}^2(q(a_{u^2}-b_{u^2})+(b-a)q_{u^2})+g_{u^1}g_{u^2}((b-a)s_{u^2}+(p-s)b_{u^2}-qb_{u^1})+rb_{u^2}g_{u^2}^2]}{(b-a) 
q g_{u^1}^2}.
\end{array}
$$
Note that  $f_{\lambda}$ and  $g_{\lambda}$ satisfy the relation
$$
(ag_{\lambda}+p+f_{\lambda})(bg_{\lambda}+s+f_{\lambda})-qr=0,
$$
which means that the covector $\theta$ is null in the conformal structure defined by the dispersion relation (\ref{dr}). 
This finishes the second proof of Theorem \ref{Thm9}.

\section{Further integrable examples in 3D}

In this section we collect miscellaneous  examples of dispersionless integrable systems in 3D related to Einstein-Weyl geometry, which do not  fit into the classes discussed above.

\medskip

\noindent {\bf Example 1.} The following system was derived by Manakov and Santini \cite{Man-San}  as a two-component generalisation of the dKP equation:
$$
u_{xt} -u_{yy} +(uu_x)_x +v_xu_{xy} -v_{y}u_{xx} =0, ~~~ v_{xt} -v_{yy} +uv_{xx} +v_xv_{xy} -v_yv_{xx} =0.
$$
Its formal linearization results upon setting $u \to u+\epsilon u_1, \ v \to v+\epsilon v_1$ which gives
$$
Lu_1+\dots=0, ~~~ Lv_1+\dots=0,
$$
here $L=\partial_x \partial_t-\partial^2_y+(u-v_y)\partial^2_x+v_x\partial_x\partial_y$, and dots denote terms which do not contain second order derivatives of $u_1$ and $v_1$. The symbol of $L$ gives rise to the conformal structure
$$
g = (dy - v_x dt )^2 - 4(dx - (u - v_y ) dt ) dt,
$$
which satisfies the Einstein-Weyl equations with the corresponding covector $\omega$ given by the formula (\ref{omega}):
$$
\omega= -v_{xx} dy+(4u_x -2v_{xy} +v_xv_{xx})dt.
$$
This Einstein-Weyl structure was obtained in \cite{Dun3}.

\medskip

\noindent {\bf Example 2.} The following system was proposed by Bogdanov \cite{Bogdanov}  as a two-component generalisation of the BF equation:
$$
(e^{-\phi})_{tt}=m_t\phi_{xy}-m_x\phi_{yt}, ~~~ m_{tt}e^{-\phi}=m_xm_{yt}-m_tm_{xy}.
$$
Its formal linearization results upon setting $\phi \to \phi+\epsilon \phi_1, \ m \to m+\epsilon m_1$ which gives
$$
L\phi_1+\dots=0, ~~~ Lm_1+\dots=0,
$$
here $L=e^{-\phi}\partial_t^2-m_x\partial_y\partial_t+m_t\partial_x\partial_y$, and dots denote terms which do not contain second order derivatives of $\phi_1$ and $m_1$. The symbol of $L$ gives rise to the conformal structure
$$
g=(m_xdx+m_tdt)^2+4e^{-\phi}m_tdxdy,
$$
which satisfies the Einstein-Weyl equations with the corresponding covector $\omega$ given by the formula (\ref{omega}):
$$
\omega=\left(\frac{m_{tt}}{m_t^{2}}-2\frac{\phi_t}{m_t}\right)(m_x\,dx+m_t\,dt)+ 2\frac{m_{yt}}{m_t}\,dy.
 $$

\medskip

\noindent {\bf Example 3.} The equation 
\begin{equation}
m_t^{\rho}m_{tt}=m_xm_{yt}-m_tm_{xy}
\label{m}
\end{equation}
was obtained in \cite{Bogdanov} as a reduction of the two-component BF system from Example 2: set $\phi=-\rho \ln m_t$. Since this equation fits into the class of quasilinear wave equations discussed in Sect. 4, it gives rise to the Einstein-Weyl geometry with the corresponding conformal structure
$$
g= (m_xdx+m_tdt)^2+4m_t^{\rho+1}dxdy,
$$
and the covector $\omega$ given by (\ref{omega}):
$$
\omega=(2\rho+1)\frac{m_{tt}}{m_t^2}(m_x\,dx+m_t\,dt)+ 2\frac{m_{yt}}{m_t}\,dy.
 $$  
Note that these $g$ and $\omega$ result, via the  substitution $\phi=-\rho \ln m_t$, from the  formulae of Example 2. It was pointed out in \cite{Bogdanov1} that, applying to this equation the contact transformation defined as
$$
(x, y, t, m, m_x, m_y, m_t)\to (x, y, \tau, F, F_x, F_y, F_{\tau})
$$
where
$$
\tau=\frac{1}{2}\ln m_t, ~~~ F=\frac{m-tm_t}{\sqrt {m_t}}, ~~~ F_x=\frac{m_x}{\sqrt{m_t}}, ~~~ F_y=\frac{m_y}{\sqrt{m_t}}, ~~~ F_{\tau}=-2t\sqrt{m_t}-\frac{m-tm_t}{\sqrt {m_t}},
$$
one obtains the equation  derived by Dunajski and Tod in the context hyper-K\"ahler metrics with conformal symmetry \cite{Dun6},
\begin{equation}
(F_{y}+F_{y\tau})(F_{ x}-F_{x \tau})-(F-F_{\tau \tau})F_{xy}=4e^{2\rho \tau}.
\label{F}
\end{equation}
It was shown in \cite{Dun6} that this equation  gives rise to the Einstein-Weyl structure
$$
g=(Fd\tau+F_xdx-F_{y}dy-dF_{\tau})^2+16e^{2\rho \tau}dxdy,
$$
$$
\omega=4\rho d\tau+\frac{(2+4\rho)(F_x-F_{x\tau})dx+(2-4\rho)(F_{y}+F_{y\tau})dy}{F-F_{\tau \tau}}.
$$
One can verify that the Einstein-Weyl structure of the equation (\ref{F}) can be obtained from that of the equation (\ref{m}) by applying the above contact transformation (and an appropriate rescaling). Note that  in the last case the covector $\omega$ is no longer given by our formula (\ref{omega}) (unless $\rho=0$, in which case  the equation becomes translationally invariant). The reason for this is that, although the Einstein-Weyl property is clearly contact-invariant, this is not the case for our formula  for $\omega$. We recall that, given a conformal structure $g$, the problem of reconstruction of the corresponding covector $\omega$ from the Einstein-Weyl constraints is far from trivial \cite{Eastwood}. What simplifies this problem  in our case is that $\omega$ should be given by a `universal' formula depending on finite order jets of a solution $u$. In any case, a fully contact-invariant approach to dispersionless integrable systems in 3D is  yet to be developed.

\section{Integrability in 4D and self-duality}

There are very few classification results  in 4D. Here we consider  the case of symplectic Monge-Amp\`ere equations, that is, equations represented as linear combinations of minors of the Hessian matrix of a function $u(x^1, x^2, x^3, x^4)$, which constitute a subclass of   equations of type III (in 4D). Below $u_{ij}$ denotes $u_{x^ix^j}$.

\bigskip

\noindent{\bf Theorem 8} \cite{DF} {\it Over the field of complex numbers, any integrable
non-degenerate symplectic  Monge--Amp\`ere equation is $Sp(8)$-equivalent to one
of the following  normal forms:
\begin{enumerate}
\item $u_{11}-u_{22}-u_{33}-u_{44}=0$ (linear wave equation);
\item $u_{13}+u_{24}+u_{11}u_{22}-u_{12}^2=0$ (second heavenly equation \cite{Plebanski});
\item $u_{13} = u_{12}u_{44}-u_{14}u_{24}$ (modified heavenly equation);
\item $u_{13}u_{24}-u_{14}u_{23}=1$ (first heavenly equation \cite{Plebanski}).
\item $u_{11}+u_{22}+u_{13}u_{24}-u_{14}u_{23}=0$ (Husain equation \cite{Husain});
\item $\alpha u_{12}u_{34} + \beta u_{13}u_{24}+\gamma u_{14}u_{23} = 0$ (general
heavenly equation),  $\alpha+\beta+\gamma=0$.
\end{enumerate}
}

\medskip

These heavenly-type equations are known to possess Lax pairs of the form  $[X_1, X_2]=0$ where $X_1, X_2$ are parameter-dependent vector fields which commute modulo the corresponding equations \cite{Plebanski, Husain, DF, Malykh}:

\medskip

\noindent {\bf Second heavenly equation}
$u_{13}+u_{24}+u_{11}u_{22}-u_{12}^2=0$:
$$
X_1=\partial_4+u_{11}\partial_2-u_{12}\partial_1+\lambda \partial_1, ~~~
X_2=\partial_3-u_{12}\partial_2+u_{22}\partial_1-\lambda \partial_2.
$$
\noindent {\bf Modified heavenly equation} $u_{13} =
u_{12}u_{44}-u_{14}u_{24}$:
$$
X_1=u_{14}\partial_2-u_{12}\partial_4+\lambda \partial_1, ~~~
X_2=-\partial_3+u_{44}\partial_2-u_{24}\partial_4+\lambda \partial_4.
$$
\noindent {\bf First heavenly equation} $u_{13}u_{24}-u_{14}u_{23}=1$:
$$
X_1=u_{13}\partial_4-u_{14}\partial_3+\lambda \partial_1, ~~~
X_2=-u_{23}\partial_4+u_{24}\partial_3-\lambda \partial_2.
$$

\noindent {\bf Husain equation} $u_{11}+u_{22}+u_{13}u_{24}-u_{14}u_{23}=0$:
$$
X_1=\partial_2+u_{13}\partial_4-u_{14}\partial_n3+\lambda \partial_1, ~~~
X_2=\partial_1-u_{23}\partial_n4+u_{24}\partial_3-\lambda \partial_2.
$$

\noindent {\bf General heavenly equation} $\alpha u_{12}u_{34} + \beta
u_{13}u_{24}+\gamma u_{14}u_{23} = 0$:
$$
X_1=\partial_1-\frac{u_{13}}{u_{34}}\partial_4+\gamma \lambda
(\partial_1-\frac{u_{14}}{u_{34}}\partial_3),  ~~~
X_2=\frac{u_{23}}{u_{34}}\partial_4-\partial_2+\beta \lambda
(\partial_2-\frac{u_{24}}{u_{34}}\partial_3).
$$

\medskip

\noindent   One can show that the  distribution spanned by $X_1$ and $X_2$ is totally null with respect to the conformal structure provided by the symbol of formal linearization.  The integrability of this distribution for any value of the spectral parameter $\lambda$ implies the existence of a three-parameter family of null surfaces ($\alpha$-surfaces), the property known to be equivalent to self-duality. Another way to see this is to notice that   the above Lax pairs are linear in  $\lambda$:
$$
X_1=X+\lambda Y, ~~~ X_2=P+\lambda Q.
$$
Furthermore, the corresponding formal linearizations can be written in the form
$$
(XQ-YP)v+\dots=0
$$
where the expression $XQ-YP$ is understood as a second order differential operator, and dots denote terms containing first order derivatives of $v$. According to \cite{Dun5, Grant} this means that the symbols of formal linearizations must be self-dual (in fact, hypercomplex, but this property is known to imply self-duality). These and other  examples support  the following conjectures relating the linearizability/integrability of four-dimensional PDEs to conformal geometry of symbols of their formal linearizations:

\begin{itemize}

\item A 4D second order dispersionless PDE is linearizable  (by a transformation from the appropriate equivalence group) if and only if  the corresponding conformal structure is conformally flat on every solution.

\item A 4D second order dispersionless PDE is integrable by the method of hydrodynamic reductions if and only if  the corresponding conformal structure is (anti) self-dual on every solution. Since the equations of self-duality are known to be integrable by the twistor construction \cite{Penrose}, this supports the evidence that solutions to integrable PDEs must carry integrable geometry.

\end{itemize}

\noindent  A convenient  approach to the integrability of four-dimensional PDEs is based on the requirement that all their travelling wave reductions to 3D must be integrable. This necessary condition turns out to be very strong indeed,
and in many cases is already sufficient for  integrability \cite{DF}. Since the integrability in 3D is related to the Einstein-Weyl condition, this gives yet another confirmation of the well-known fact that symmetry reductions of the self-duality equations lead to  Einstein-Weyl geometry \cite{ Jones, Calderbank, Calderbank1}.

\section{Appendix: the method of hydrodynamic reductions}

As proposed in \cite{Fer1}, the method of hydrodynamic reductions applies to first order quasilinear systems of the  form
\begin{equation}
A ({\bf u}){\bf u}_x+B({\bf u}){\bf u}_y+C({\bf u}){\bf u}_t=0,
\label{quasi}
\end{equation}
or equations transformable into this form by a suitable change of variables. Here ${\bf u}=(u^1, ..., u^m)^t$ is an $m$-component column vector of the dependent variables, and $A, B, C$ are $l\times m$ matrices where $l$, the number of equations, is allowed to exceed the number of the unknowns, $m$. The system (\ref{quasi}) will   be assumed involutive with the general solution depending on a certain number of arbitrary functions of {\it two} variables. Systems of this type are referred to as 3D  dispersionless PDEs.  Typically, they arise as dispersionless limits of integrable soliton equations: the canonical example is the KP equation, 
$u_{t}-uu_x +\epsilon^2 u_{xxx}-w_{y}=0, \ w_x=u_y,$ which assumes the form (\ref{quasi})  in the   limit $\epsilon \to 0$. 

It will be demonstrated below that equations of types  (\ref{T1})--(\ref{T3}) are indeed within the class (\ref{quasi}).
The method of hydrodynamic reductions consists of seeking $N$-phase solutions in the form
\begin{equation}
{\bf u}={\bf u}(R^1, ..., R^N)
\label{phase}
\end{equation}
where the `phases' $R^i(x, y, t)$ are required to satisfy a pair of consistent equations
\begin{equation}
 R^i_y=\mu^i(R) R^i_x,  ~~~~ R^i_t=\lambda^i(R) R^i_x,
\label{R}
\end{equation}
$i=1,\dots, N$, no summation.  The variables $R^i$ are also known as {\it Riemann invariants}.  We recall that the compatibility conditions, $R^i_{yt}=R^i_{ty}$, 
imply the following restrictions for the
characteristic speeds $\mu^i$ and $\lambda^i$:
\begin{equation}
\frac{\partial_j\mu^i}{\mu^j-\mu^i}=\frac{\partial_j\lambda
^i}{\lambda^j-\lambda^i}, 
\label{comm}
\end{equation}
$i\ne j, ~ \partial_i=\partial/\partial_{R^i}$. Commuting systems of the form (\ref{R}) can be solved by the generalized hodograph method  \cite{Tsarev}. Equations of the form (\ref{R}) are known as $N$-component {\it systems of hydrodynamic type}, their  Hamiltonian and geometric aspects have been thoroughly investigated in \cite{Dub, Tsarev}. 
We require that {\it every} solution of (\ref{R}) gives rise to a solution of the original system (\ref{quasi}) via (\ref{phase}). In this case equations (\ref{R}) are said to constitute an $N$-component  {\it hydrodynamic reduction} of the original system. Each hydrodynamic reduction can be viewed as a decomposition of the original $m$-component 3D system into a pair of commuting $N$-component 2D systems. 
 It is remarkable that there exist 3D systems possessing an infinity of  such reductions:

\medskip

\noindent {\bf Definition. }  \cite{Fer1}
{\it The system (\ref{quasi}) is said to be {\it integrable by the method of hydrodynamic reductions} if, for any $N$,  it possesses infinitely many $N$-component hydrodynamic reductions   parametrized by $N$ arbitrary functions of one variable}. 

\medskip

One can show that the existence of $3$-component reductions depending on $3$ arbitrary functions of one variable is already sufficiently restrictive, and implies the integrability. This is reminiscent of the well-known 3-soliton condition in the Hirota bilinear approach. For a particular  reduction, the corresponding solutions constitute only a very narrow subclass of solutions of the original system (\ref{quasi}). As $N$ varies, solutions coming from $N$-component hydrodynamic reductions form an ever growing subset of solutions of the system (\ref{quasi}) which is, in a sense,  locally dense (that is, these solutions do not satisfy any finite order differential constraints other than the equation itself, and its differential consequences). Multi-phase solutions of this form originate from gas dynamics \cite{Sidorov, Burnat, Perad, Grundland}, see also references therein. They are sometimes referred to as {\it nonlinear interactions of planar simple waves}. For $N=1,2$ they are called {\it simple waves} and {\it double waves}, respectively, and belong  to the class of  solutions with a degenerate hodograph.

The above definition provides an efficient classification criterion. In general, one proceeds as follows: substituting (\ref{phase}) into (\ref{quasi}) and using (\ref{R}) one obtains
\begin{equation}
(A+\mu^iB+\lambda^iC)\partial_i{\bf u}=0.
\label{ABC}
\end{equation}
The condition of the nontrivial solvability of this linear system provides the dispersion relation for the characteristic speeds $\mu^i$ and $\lambda^i$ for any $i=1, \dots, N$. Namely, the pair $\lambda, \mu$
(more precisely, the covector $dx+\mu dy+\lambda dt$) is called characteristic if
$$
rk(A+\mu B+\lambda C)<m.
$$
We will assume that the dispersion relation defines an irreducible algebraic curve. For instance, in the case $l=m$ this gives an algebraic curve of degree $m$,
$$
det(A+\mu B+\lambda C)=0.
$$
In the language of PDEs this is the affine part of the characteristic variety of our system \cite{Kras}.
For all examples discussed in this paper, the dispersion relation reduces to a non-degenerate conic determined by the symbol of formal linearization. Equations (\ref{comm}) and (\ref{ABC}) form an overdetermined system for the functions ${\bf u}(R)$ and the characteristic speeds  $\mu^i(R), \lambda^i(R)$. The requirement that they are consistent, with the general solution depending on $N$ arbitrary functions of one variable, leads to the integrability conditions in terms of the original matrices $A, B, C$. Let us illustrate the method of hydrodynamic reductions using the example of the dKP equation, in which case all calculations can be  verified by hand. 

\medskip

\noindent {\bf Example 1: hydrodynamic reductions of the dKP equation.} The  dKP equation,
$
u_{xt}-(uu_{x})_x-u_{yy}=0,
$
can be written in the   form (\ref{quasi}):
\begin{equation}
u_t-uu_x=w_y, ~~~ u_y=w_x.
\label{dKP}
\end{equation}
Looking for $N$-phase solutions,  $u=u(R^1, ..., R^N), \ w=w(R^1, ..., R^N)$,
where the phases $R^i$ satisfy Eqs. (\ref{R}), one  obtains the relations
\begin{equation}
\partial_i w=\mu^i \partial_i u,  ~~~  \lambda^i=u+(\mu^i)^2,
\label{110}
\end{equation}
(the second one is the dispersion relation). The compatibility condition
$\partial_i\partial_jw=\partial_j\partial_iw$  implies
\begin{equation}
\partial_i\partial_ju=\frac{\partial_j\mu^i}{\mu^j-\mu^i}\partial_iu+\frac{\partial_i\mu^j}{\mu^i-\mu^j}\partial_ju,
\label{11}
\end{equation}
which, along with  the commutativity conditions (\ref{comm}), result in the following system for $u(R)$ and $\mu^i(R)$, known as the Gibbons-Tsarev system,
\begin{equation}
\partial_j\mu^i=\frac{\partial_j u}{\mu^j-\mu^i}, ~~~
\partial_i\partial_ju=2\frac{\partial_iu\partial_ju}{(\mu^j-\mu^i)^2},
\label{13}
\end{equation}
$i, j=1, \dots, N, ~ i\ne j$, which was first derived in \cite{GibTsar} in the
context of hydrodynamic reductions of the
Benney moment equations, see also \cite{Gib94}.
For any solution $\mu^i,  u$ of the system (\ref{13}) one can
reconstruct $\lambda^i$ and $w$ by virtue of (\ref{110}).
The system (\ref{13}) is compatible and its general solution 
depends, modulo transformations $R^i\to f^i(R^i)$, on $N$
arbitrary functions of one variable. This gives the required family of $N$-component hydrodynamic reductions parametrized by $N$ arbitrary functions of one variable, and establishes the integrability of dKP. We point out that the compatibility conditions,  $\partial_k\partial_j\mu^i=\partial_j\partial_k\mu^i$ and $\partial_k\partial_i\partial_ju=\partial_j\partial_i\partial_ku$, involve {\it triples} of indices $i\ne j \ne k$ only. Thus, the consistency of the system (\ref{13}) for $N=3$ implies 
its consistency for arbitrary $N$. This turns out to be a general phenomenon: as mentioned above, the existence of 3-component reductions is already sufficient for the integrability. 

Let us demonstrate that equations  (\ref{T1})--(\ref{T4}) can be brought into the first order quasilinear form (\ref{quasi}). 

\medskip

\noindent {\bf Equations of type I:}
$$
a_{xx} + b_{yy} + c_{tt} + 2 p_{xy} + 2 q_{xt} + 2 r_{yt} = 0,
$$
recall that $a, b, c, p, q, r$ are functions of one and the same dependent variable $u$. Consider the first order system
$$
a_x+p_y+q_t=-\varphi_t, ~~~ b_y+p_x+r_t=\psi_t, ~~~ c_t+q_x+r_y=\varphi_x-\psi_y,
$$
which implies the above second order PDE on elimination of the auxiliary potentials $\varphi, \psi$. This system is invariant under  gauge transformations $\varphi \to \varphi+\eta_y, \ \psi \to \psi+\eta_x$ where $\eta$ is an arbitrary function of $x, y$. One can show that this gauge freedom can be eliminated by imposing a compatible differential constraint
$$
a'\psi_x+ p'\psi_y+q'\psi_t+p'\varphi_x+b'\varphi_y+r'\varphi_t=0.
$$
This results in a system of  the form (\ref{quasi}) with ${\bf u}=(u, \varphi, \psi)$ and $(l, m)=(4, 3)$. It was investigated by the method of hydrodynamic reductions in \cite{Fer5} (see Theorem \ref{Thm1} for the integrability conditions).

\smallskip

\noindent {\bf Equations of type II:}
$$
f_{11} u_{xx} + f_{22} u_{yy} + f_{33} u_{tt} + 2 f_{12} u_{xy} + 2f_{13} u_{xt} + 2f_{23} u_{yt} =0,
$$
recall that $f_{ij}$ are functions of the first order derivatives of $u$ only. Setting $u_x=a, \ u_y=b, \ u_t=c$, one obtains an equivalent  first order representation,
$$
a_y=b_x, ~~~ a_t=c_x, ~~~ b_t=c_y, ~~~ 
f_{11} a_x+ f_{22} b_y + f_{33} c_t+ 2f_{12} a_y  + 2f_{13} a_t  + 2f_{23} b_t=0,
$$
where $f_{ij}$ are now viewed as functions of $a, b, c$. This is again of the form (\ref{quasi}) with ${\bf u}=(a, b, c)$ and $(l, m)=(4, 3)$. This class was investigated by the method of hydrodynamic reductions in \cite{Bur}. 

\smallskip

\noindent {\bf Equations of type III:}
$$
F(u_{xx}, u_{xy}, u_{yy}, u_{xt}, u_{yt}, u_{tt})=0.
$$
Rewriting the equation in explicit form,
$$
u_{tt}=f(u_{xx}, u_{xy}, u_{yy}, u_{xt}, u_{yt}),
$$
and setting
$
u_{xx}=a,\ u_{xy}=b, \ u_{yy}=c,\ u_{xt}=p,\ u_{yt}=q, \ u_{tt}= f(a,b, c, p, q),
$
one obtains a first order quasilinear system by writing out all possible consistency conditions among the second order derivatives,
$$
\begin{array}{c}
a_y=b_x, ~~ a_t=p_x, ~~ b_y=c_x, ~~ b_t=p_y=q_x, ~~ c_t=q_y, \\
\ \\
 p_t=f(a,b, c, p, q)_x, ~~ q_t=f(a,b, c, p, q)_y.
\end{array}
$$
This is of the form (\ref{quasi}) with ${\bf u}=(a, b, c, p, q)$ and $(l, m)=(8, 5)$. This class was investigated by the method of hydrodynamic reductions in \cite{Fer4}.

\smallskip

\noindent {\bf Equations of type IV:}
$$
A(u)u_x+B(u)u_y+C(u)u_t=0.
$$
This system is of the form (\ref{quasi}), with $(l, m)=(2, 2)$. This class was investigated by the method of hydrodynamic reductions in \cite{FK}. 

\bigskip

\noindent {\bf Remark 1: hydrodynamic reductions of elliptic PDEs.} Although the method of hydrodynamic reductions  was primarily designed to deal with hyperbolic systems, it can be extended to elliptic PDEs. The only difference is that, in the elliptic case, the dispersion curve (characteristic variety) has no real points, so that the characteristic speeds $\mu^i$, $\lambda^i$, as well as the variables $R^i$, must be complex.  We will say that an elliptic system (\ref{quasi}) with real-analytic coefficients is integrable by the method of hydrodynamic reductions if it possesses `infinitely many' complex-valued solutions (\ref{phase}) where the complex 
phases $R^i$ satisfy the commuting systems (\ref{R}). In this context,  ${\bf u}, \mu^i, \lambda^i$ are viewed as complex-analytic functions of $R^i$. We would like to stress that the integrability conditions for elliptic PDEs, as well as the procedure of  their derivation, are  identically the same as in the hyperbolic case. Since the integrability conditions is all we need to prove the Einstein-Weyl property of the symbol of formal linearization, it will still hold for elliptic integrable systems. 
On imposing appropriate `reality' constraints, the above scheme can  lead to elliptic reductions with {\it real} coefficients. Let us demonstrate this using again the example of dKP. 

\medskip

\noindent {\bf Example 2: elliptic reduction of the dKP equation.} Two-phase solutions of the system 
(\ref{dKP}) have  the form  $u=u(R^1, R^2), \ w=w(R^1, R^2)$ 
where the phases $R^1, R^2$ satisfy Eqs. (\ref{R}).  As shown in Example 1, one  obtains the relations (\ref{110}) for $w$ and $\lambda^i$, along with the Gibbons-Tsarev equations (\ref{13}) for $\mu^i$ and $u$:
$$
\partial_2\mu^1=\frac{\partial_2 u}{\mu^2-\mu^1}, ~~~ \partial_1\mu^2=\frac{\partial_1 u}{\mu^1-\mu^2}, ~~~
\partial_1\partial_2u=2\frac{\partial_1u\partial_2u}{(\mu^2-\mu^1)^2}.
$$
A particular solution of these equations (the so-called shallow water reduction) is given by
$$
\mu^1=R^2+3R^1, ~~~ \mu^2=R^1+3R^2, ~~~ u=(R^1-R^2)^2,
$$
so that relations (\ref{110}) imply
$$
w=2(R^1-R^2)^2(R^1+R^2), ~~~ \lambda^1=2(R^1+R^2)^2+8(R^1)^2,  ~~~ \lambda^2=2(R^1+R^2)^2+8(R^2)^2.
$$
Let us now allow $R^1$ and $R^2$ to be complex, such that $R^2=\bar R^1$. Setting 
$R^1=p+iq, \ R^2=p-iq$ we obtain $u=-4q^2$, $w=-16pq^2$ (both real!), while equations (\ref{R}) result, on separating real and imaginary parts, in the following pair of two-component commuting elliptic systems with real coefficients:
$$
\left(\begin{array}{c}
p \\
q
\end{array}\right)_y=
\left(\begin{array}{cc}
4p & -2q\\
2q & 4p
\end{array}\right)
\left(\begin{array}{c}
p \\
q
\end{array}\right)_x, ~~~
\left(\begin{array}{c}
p \\
q
\end{array}\right)_t=
\left(\begin{array}{cc}
16p^2-8q^2 & -16pq\\
16pq &16p^2-8q^2
\end{array}\right)
\left(\begin{array}{c}
p \\
q
\end{array}\right)_x.
$$
Thus we obtain a two-component elliptic reduction of dKP. 

\bigskip

\noindent {\bf Remark 2:  hydrodynamic reductions in higher dimensions.} The method of hydrodynamic reductions generalizes to higher dimensional quasilinear systems, say 
$$
A ({\bf u}){\bf u}_x+B({\bf u}){\bf u}_y+C({\bf u}){\bf u}_z+D({\bf u}){\bf u}_t=0.
$$
One again looks at $N$-phase solutions in the form
$$
{\bf u}={\bf u}(R^1, ..., R^N),
$$
where the phases $R^i(x, y, t)$  satisfy a triple of consistent equations
$$
 R^i_y=\mu^i(R) R^i_x,  ~~~~ R^i_t=\lambda^i(R) R^i_x, ~~~~ R^i_z=\eta^i(R) R^i_x,
$$
see \cite{FerPav, FKmulti} for further details.

\section{Concluding remarks}

\noindent {\bf (a)} We have characterized the integrability of several classes of  dispersionless PDEs in 3D by the Einstein-Weyl property of their formal linearizations. It should be  emphasized that the four types of equations under consideration, although invariant under certain  equivalence groups,  are clearly not contact/point invariant.
On the other hand, the property for a second order PDE to have the Einstein-Weyl symbol of formal linearization, is invariant under arbitrary contact (and more generally Lie-B\"acklund type) transformations. 
Although the covector $\omega$ will no longer be given by the simple formula (\ref{omega}), it will still depend on finite order jets of the transformed equation. 
This suggests  a contact-invariant approach to the dispersionless integrability  unifying all known examples. The class of Monge-Amp\`ere equations in 3D would be the natural venue to develop a fully contact-invariant theory. 
We should however warn the reader that, in general,  the Einstein-Weyl property alone may not be sufficient  for the dispersionless integrability: as we saw in the proof of Theorem~\ref{Thm8}, the system $u^1_t=qu^2_y, \  u^1_y={q}u^2_x$ has a conformally flat symbol, but  is not integrable for generic $q(u^1, u^2)$. 



\medskip

\noindent {\bf (b)} Solutions of dispersionless PDEs discussed in this paper, and the induced Einstein-Weyl structures, are generally defined on open domains in $\bbbr^3$. One cannot guarantee more since the construction applies to {\it  all} solutions. We believe however that, for some particular solutions, global aspects/singularities may come into play. As an illustration let us consider equations (\ref{T3}) of the dispersionless Hirota type. Geometrically, an equation of this form  specifies a hypersurface $M^5$ in the Lagrangian Grassmannian $\Lambda^6$ \cite{Fer4}. For some particular integrable equations, such as e.g. the potential dKP equation $u_{xt}-u_{xx}^2-u_{yy}=0$, this hypersurface is algebraic. Solutions of the equation correspond to Lagrangian submanifolds in the six-dimensional symplectic space whose Gaussian image belongs to $M^5$. 
Some of them may give rise to algebraic 3-folds in $M^5$, with nontrivial global properties. The investigation of  algebraic solutions and the induced Einstein-Weyl structures is one of the interesting problems left outside the scope of this paper.

\medskip

\noindent {\bf (c)} In the case of higher order dispersionless PDEs in 3D, the symbol of formal linearization defines a generalized conformal structure which supplies each solution with a field of algebraic null cones. In the spirit of \cite{Cartan}, one can define the `generalized Einstein-Weyl' property by the requirement of the existence of a symmetric connection which preserves this generalized conformal structure, and possesses  a two-parameter family of null totally geodesic surfaces. We expect that our results will carry over to this more general context.

\medskip

\noindent {\bf (d)} Attempts to extend our results to dimensions higher than four meet an immediate obstacle: all known integrable (non-linearizable) examples in dimensions five and higher have degenerate symbols.   Thus, their solutions do not carry any conventional `geometry'. One possible way to proceed is to require that  non-degenerate travelling wave reductions of such PDEs to 3D/4D give rise to Einstein-Weyl/self-dual geometries. 

\medskip

\noindent {\bf (e)} Einstein-Weyl structures are known to be related to third order ODEs satisfying the so-called W\"unschmann and Cartan relations \cite{Cartan1, Tod1, Nurowski}. Thus, every solution of a dispersionless integrable system has a third order ODE naturally associated with it (to be precise, a point equivalence class of ODEs).  The (complexified) space of the dependent and independent variables of this ODE, which can be identified with the space of null totally geodesic  surfaces of the Einstein-Weyl structure, is known as the `minitwistor' space \cite{Hitchin, Tod}.  Solutions of this ODE correspond to curves of   totally geodesic surfaces passing through a fixed point: these  form a three-parameter family, and are known as `twistor curves'. It would be of interest to develop an efficient procedure to explicitly calculate this ODE for  multi-phase solutions  provided by the method of hydrodynamic reductions.

\section*{Acknowledgements}

We thank D. Calderbank,
 M. Dunajski,  K. Khusnutdinova and M. Pavlov
for useful comments and clarifying discussions. We also thank the referee for numerous suggestions which helped to improve  the presentation.  
We acknowledge financial support from the LMS (BK) and  the University of Troms\o\ (EVF)  
making this collaboration possible.
Special thanks go to the University Hospital of North Norway where part of this work was completed.
The research of EVF was partially supported  by the European Research Council Advanced Grant FroM-PDE.

\end{document}